\documentclass{aa} 
\usepackage{graphicx,url,twoopt,natbib}
\usepackage{hyperref}                %% for pdflatex
\hypersetup{ colorlinks=true,   urlcolor=blue,  citecolor=blue,   linkcolor=magenta}
\usepackage{graphics}
\usepackage{txfonts}
\usepackage{graphicx}
\usepackage{amsmath}    % Advanced maths commands
\usepackage{amssymb}    % Extra maths symbols
\usepackage{bm}     % Bold maths symbols, including upright Greek
\usepackage{pdflscape}  % Landscape pages
\usepackage[ pagewise,  switch]{lineno}%%  line number
\usepackage{xcolor}
\usepackage{ae,aecompl}
\usepackage{subfigure}
\usepackage{scalerel}
\usepackage[english]{babel}
\usepackage{times}
\usepackage{physics}
\usepackage{etoolbox}
\usepackage{multirow, booktabs}
\usepackage{siunitx}
\usepackage{lipsum}
\usepackage{paralist}
\usepackage{makecell}
\usepackage{tabularx}
\usepackage[inline]{enumitem}
\newlist{mycompactenum}{enumerate}{1}
\setlist[mycompactenum,1]{nosep,label=\arabic*.}
\usepackage{paralist}
 \bibpunct{(}{)}{;}{a}{}{,} % to follow the A&A style
\makeatletter
\protected\def\stonyslink{%
	\def\hyper@linkstart##1##2{}\let\hyper@linkend\@empty}
\newcommandtwoopt{\citeads}[3][][]{%
	\href{http://ui.adsabs.harvard.edu/abs/#3/abstract}%
	{\stonyslink \citealp[#1][#2]{#3}}%   %% Rutten, 2000
	\biblink{#3}{\href{http://ui.adsabs.harvard.edu/abs/#3/abstract}{ADS}}}
\newcommandtwoopt{\citepads}[3][][]{%
	\href{http://ui.adsabs.harvard.edu/abs/#3/abstract}%
	{\stonyslink \citep[#1][#2]{#3}}%     %% (Rutten 2000)
	\biblink{#3}{\href{http://ui.adsabs.harvard.edu/abs/#3/abstract}{ADS}}}
\newcommandtwoopt{\citetads}[3][][]{%
	\href{http://ui.adsabs.harvard.edu/abs/#3/abstract}%
	{\stonyslink \citet[#1][#2]{#3}}%     %% Rutten (2000)
	\biblink{#3}{\href{http://ui.adsabs.harvard.edu/abs/#3/abstract}{ADS}}}
\newcommandtwoopt{\citeyearads}[3][][]{%
	\href{http://ui.adsabs.harvard.edu/abs/#3/abstract}%
	{\stonyslink \citeyear[#1][#2]{#3}}%  %% 2000
	\biblink{#3}{\href{http://ui.adsabs.harvard.edu/abs/#3/abstract}{ADS}}}
\makeatother

\begin{document} 
%\linenumbers

\title{Variation of the stellar color in high-magnification and caustic-crossing microlensing events \thanks{Table \ref{tablast} are only available in electronic form at the CDS (\url{/CatS/188.136.164.241:002})}}

\author{S. Sajadian \inst{1, 2} \and U.~G.~J\o rgensen\inst{3}}
\institute{Department~of~Physics,~Isfahan~University~of~Technology,~Isfahan~84156-83111,~Iran\\ \email{s.sajadian@iut.ac.ir} \and Department~of~Physics,~Chungbuk~National~University,~Cheongju~28644,~Republic~of~Korea \and Centre~for~ExoLife~Sciences~(CELS),~Niels~Bohr~Institute,~University of Copenhagen,~\O stervoldgade~5,~DK-1350~Copenhagen,~Denmark\\ \email{uffegj@nbi.dk}}
 	
%\date{Received September 15, 1996; accepted March 16, 1997}
% \abstract{}{}{}{}{} 
% 5 {} token are mandatory

\abstract 
{To a first approximation, the microlensing phenomenon is achromatic, and great advancement has been achieved in the interpretation of the achromatic signals, which among other achievements has led to the discovery and characterization of well above $100$ new exoplanets. At higher order accuracy in the observations, microlensing has a chromatic component (a color term) which has so far been much less explored.}
{Here, we analyze the chromatic microlensing effect of $4$ different physical phenomena, which have the potential to add important new knowledge about the stellar properties not easily reachable with other methods of observations. Our simulation is limited to the case of main-sequence source stars.}
{Microlensing is particularly sensitive to giant and sub-giant stars near the Galactic center. While this population can be studied in short snapshots by use of the largest telescopes in the world, a general monitoring and characterization of the population can be achieved by use of more accessible medium-sized telescopes with specialized equipments through dual-color monitoring from observatories at sites with excellent seeing. We quantify our results to what will be achievable from the Danish $1.54$m telescope at La Silla observatory by use of the existing dual-color lucky imaging camera. Such potential monitoring programs of the bulge population from medium-sized telescopes include the characterization of starspots, limb-darkening, the frequency of close-in giant planet companions, and gravity darkening for blended source stars.}
{We conclude our simulations with quantifying the likelihood of detecting these different phenomena per object where they are present to be $\sim 60\%$ and $\sim 30\%$ for the mentioned phenomena, when monitored during high magnification and caustic crossings, respectively.}
 {}


\maketitle

\section{Introduction}\label{twosec}
According to Einstein's theory of general relativity, the light path of a background star (the source) is bent while passing through the gravitational field of a foreground object (the lens star, or any other massive object). The amount of this bending angle is \citep{Einstein1936}:  
\begin{eqnarray}
\theta \simeq \frac{4~G~M_{\rm l}}{c^{2}~b},
\end{eqnarray}
\noindent where $M_{\rm l}$ is the mass of the lensing object, $b$ is the impact parameter (i.e., the minimum distance of a light ray and the lensing object), $G$ is the gravitational constant and $c$ is the speed of light. Hence, in a lensing phenomenon, the bending angle does not depend on the wavelength of the light. A gravitational red and blue shift of the light does occur, because of the component of the gravitational force parallel to the light path, but it is symmetric in direction while the light approaches and moves away from the lensing object, so the wavelength before and after the passage remains the same.  

In a lensing phenomenon, two or more distorted images are formed and their angular distance depends on the physical lens and source distances from the observer and on the mass of the lens. When the lens and source stars are inside the Galactic disk and bulge respectively, the angular distance of the images is less than a milli arc second, too small to be resolved. This kind of lensing is the so-called gravitational microlensing \citep{Liebes1964, Chang1979, Pac1986}.

One important quantity in microlensing events that can be inferred from observations is the magnification factor of the source flux relative to its value at the baseline, i.e., $A(t)$. In the point-source approximation it is given by \citep{Einstein1936}:
\begin{eqnarray}
A(t)=\frac{u^2(t) +2}{u(t)~\sqrt{u^2(t)+4}}, \label{simpleA}
\end{eqnarray}
where, $u(t)$ is the angular lens-source distance normalized to the angular Einstein radius $\theta_{\rm E}$ (angular radius of the ring of the images for perfect alignment) \citep[see, e.g., ][]{book1992, gaudi2012}.

%%p4
In a more realistic treatment of the microlensing phenomenon, where the source star is considered to have a finite size rather than being a point,  each infinitesimally small area of the source star is usually treated as having the same color. This will result in a realistic evaluation of $A(t)$, but will still result in an achromatic magnification factor. This approximation has been very successful for characterizing even details of exoplanets orbiting the lensing stars. However, a real star of course does not have constant color over its surface, and during a lensing phenomenon various elements of the source star will be amplified with various strengths as a function of time, depending on their position on the source surface, and the resulting magnification light curve will therefore in reality be color-dependent (i.e., chromatic).

%%p5
These variations are small compared to the general magnification, but they contain important information about the source stars that are not easily obtainable in other ways. We have already above mentioned various such systematic color variations, and we will show in the following sections that they are measurable for medium-sized ground based telescopes. Of particular importance is the color variations due to close-in giant planets transiting the source star during the lensing. Generally, there are two main channels for detecting exoplanets through microlensing observations: (i) detecting gravitational perturbations on the source brightness due to exoplanets orbiting lens objects \citep[see, e.g.,][]{1991MAO, Gould1992, 2018Tsapras} and (ii) detecting perturbations on the source brightness profiles or their astrometric motions while lensing due to giant planets orbiting them \citep[e.g., ][]{2000Graff,2009xarallap,2019Bagheri}. 

\noindent Microlensing measured from dual-color instruments on medium-sized telescopes, such as the Danish $1.54$ m telescope at ESO's La Silla observatory in Chile \citep{2012Harpsoe,Scottfelt2015}, will to the best of our knowledge be the only viable way of obtaining statistics on the exoplanet population of stars near the Galactic center, and thereby adding important new information to solving the question about the dependence of exoplanetary frequencies on the galactic environment. Ongoing high-magnification (hereafter HM) and caustic-crossing (hereafter CC) microlensing events have been (and are) followed up by the Danish telescope since 2008 \citep{2008Dominik}. In several cases, its observations made better resolutions of the planet signatures and more precise characterizations of these planets \citep[see, e.g., ][]{2018Udalski, 2018Hanb, 2018Ryu, Li2019variable, 2019Street, 2020Hirao, 2021Konda}.

The chromatic change of limb-darkened source stars in HM microlensing events was first investigated by \citet{Valls1998}. Then, the chromatic changes of limb-darkened source stars in HM and CC microlensing events by considering the blending (BL) effect have been studied in \citet{Han2000a,Han2001ch,Pejcha2009}. The chromatic variations of highly blended source stars during microlensing events can reveal their types \citep{Tsapras2019}. Such measurements are done by multiband follow-up microlensing observations with the ROME/REA project. Recently, the chromatic deviations in HM or CC microlensing events due to radially and non-radially pulsating source stars have extensively been studied \citep{SajadianIgnaceI, sajadianIgnaceII, Sajadian2021III}. All of these studies emphasize the importance of detection and characterization of chromatic deviations during microlensing events.

In this work, we extend the previous researches by considering more realistic models for the limb-darkening (LD) effect for different types of source stars based on their atmospheric models and including the blending effect. Here, the chromatic effects are most accurately studied during HM, and CC features where the relative differences between the magnification of different parts of the source star is largest.

\noindent We therefore simulate in this paper HM and CC microlensing events toward the Baade's window of blended source stars with variable color over their surfaces, and study the resulted chromatic perturbations. Five causes for variable source star surface color studied below are the BL, LD, close-in giant planet companions (CGPs), stellar spots (SS), and finally gravity-darkening (GD) for hot stars, which are described separately in Sections \ref{blends}, \ref{limb}, \ref{CGPs}, \ref{spots} and \ref{gradark}, respectively. In the last section of the paper, \ref{result}, we summarize the conclusions.
\begin{figure}
	\centering
	\includegraphics[angle=0,width=0.49\textwidth,clip=0]{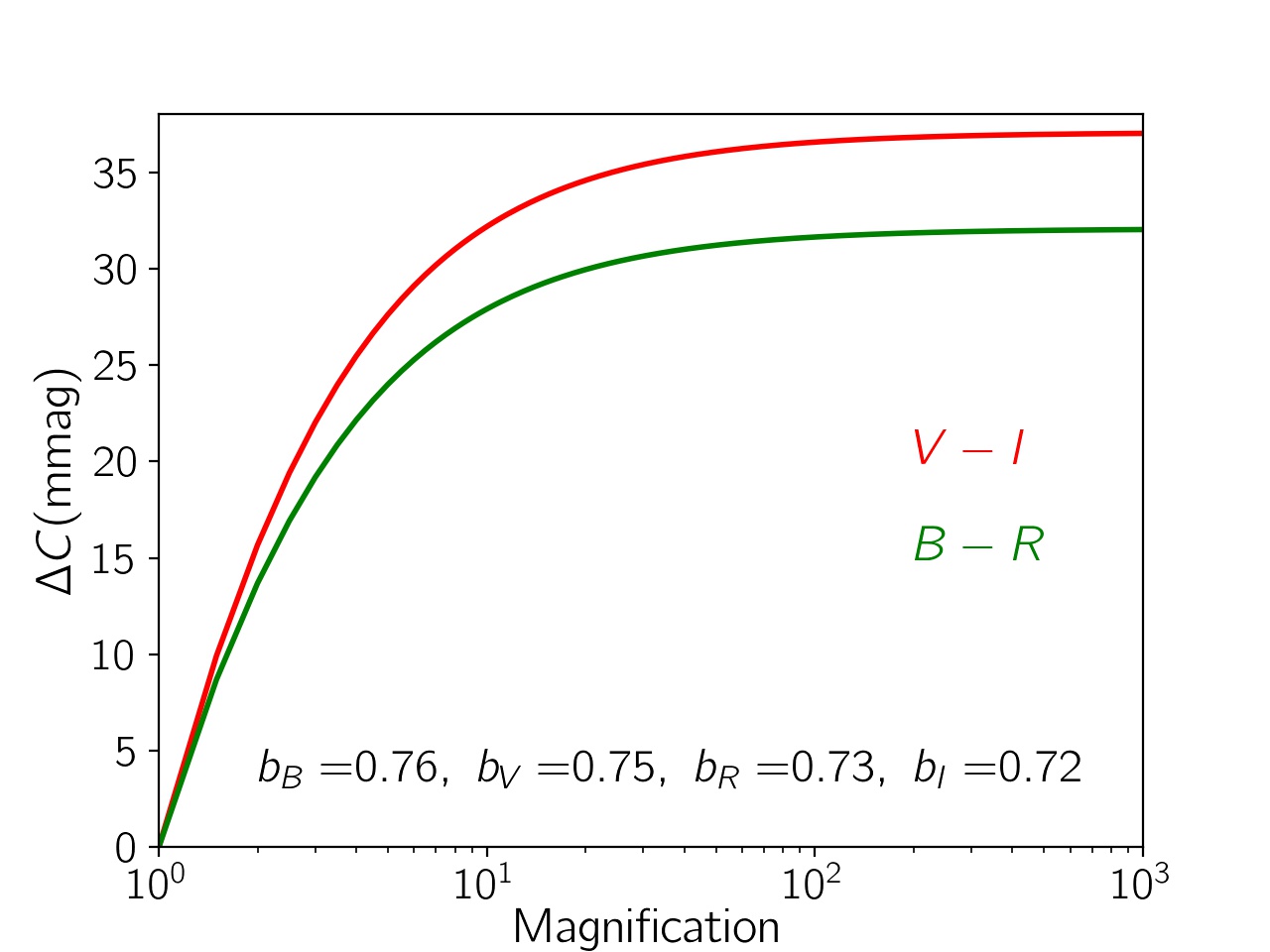}
	\caption{The variation in the stellar colors $V$-$I$ (red) and $B$-$R$ (green) versus the magnification factor due to the BL effect. The adopted blending parameters, given inside the figure, are the average of blending parameters for simulated microlensing events toward the Baade's window, with the coordinate $(l_{\rm G}=1^{\circ},~b_{\rm G}=-4^{\circ})$, by excluding highly blended events with $b_{I}<0.2$.}\label{blendd}
\end{figure}

%%%%%%%%%%%%%%%%%%%%%%%%%%%%%%%%%%%%%%%%%%%%%%%%%%%%%%%%%%%%%%%%%%%%
\section{Blending effect}\label{blends}
The microlensing observations have been (and are) mostly done toward the Galactic bulge. In these directions, the column number density of stars is high, which makes that the source light is blended with other close stars, the so-called blending effect \citep[see, e.g., ][]{1995DiStefano, 2007Smith}. To evaluate this effect, we determine the average cumulative number of stars in a given direction and distance, as follows:
\begin{eqnarray}
\left<N_{\rm{b}}(l_{\rm G}, b_{\rm G}, D_{\rm s})\right>= \Omega_{\rm{PSF}}~\int_{0}^{D_{\rm s}} dD~D^{2}~\sum_{i=1}^{4} \frac{\rho_{i}(l_{\rm G},~b_{\rm G},~D)}{\left<M_{i}\right>}, 
\end{eqnarray}
where, $(l_{\rm G},~b_{\rm G})$ specify the Galactic longitude and latitude, $D$ is the distance which changes from zero to $D_{\rm s}$ (the source position with respect to the observer), and $\Omega_{\rm{PSF}}$ is angular size of the source PSF (Point Spread Function). The summation is done over different Galactic structures including the Galactic thin and thick disks, bulge and the stellar halo. Here, $\rho_{i}$ and $\left<M_{i}\right>$ represent the stellar mass density and the average mass of these Galactic structures, respectively. Toward a given direction, the number of blending stars is $N_{\rm b}=\left<N_{\rm b}\right> + \delta N_{\rm b}$, where  $\delta N_{\rm b}$ is determined from a normal distribution with the width $\sqrt{\left<N_{\rm b}\right>}$. For each of these blending stars, we determine its flux in a given filter, $\mathcal{F}_{F}$, using the Galactic Besan\c{c}on model \citep{Robin2003,Robin2012}. The index $F$ refers to the adopted filter and $F \in UBVRI$. More explanations can be found in \citet{Moniez2017, Sajadian2019, 2021sajadian}.
\begin{figure*}
\centering
\subfigure[]{\includegraphics[width=0.48\textwidth]{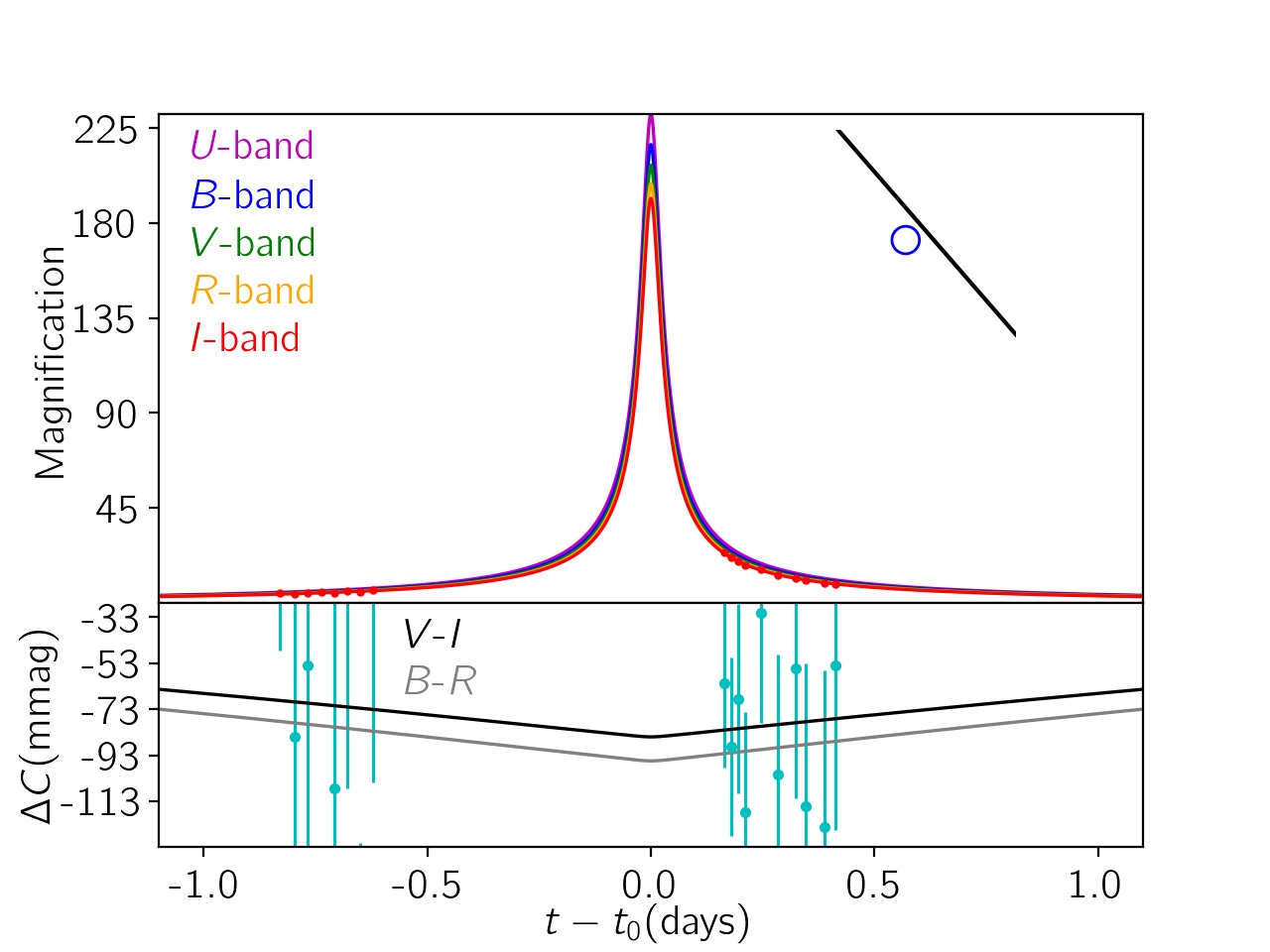}\label{fig0a}}
\subfigure[]{\includegraphics[width=0.48\textwidth]{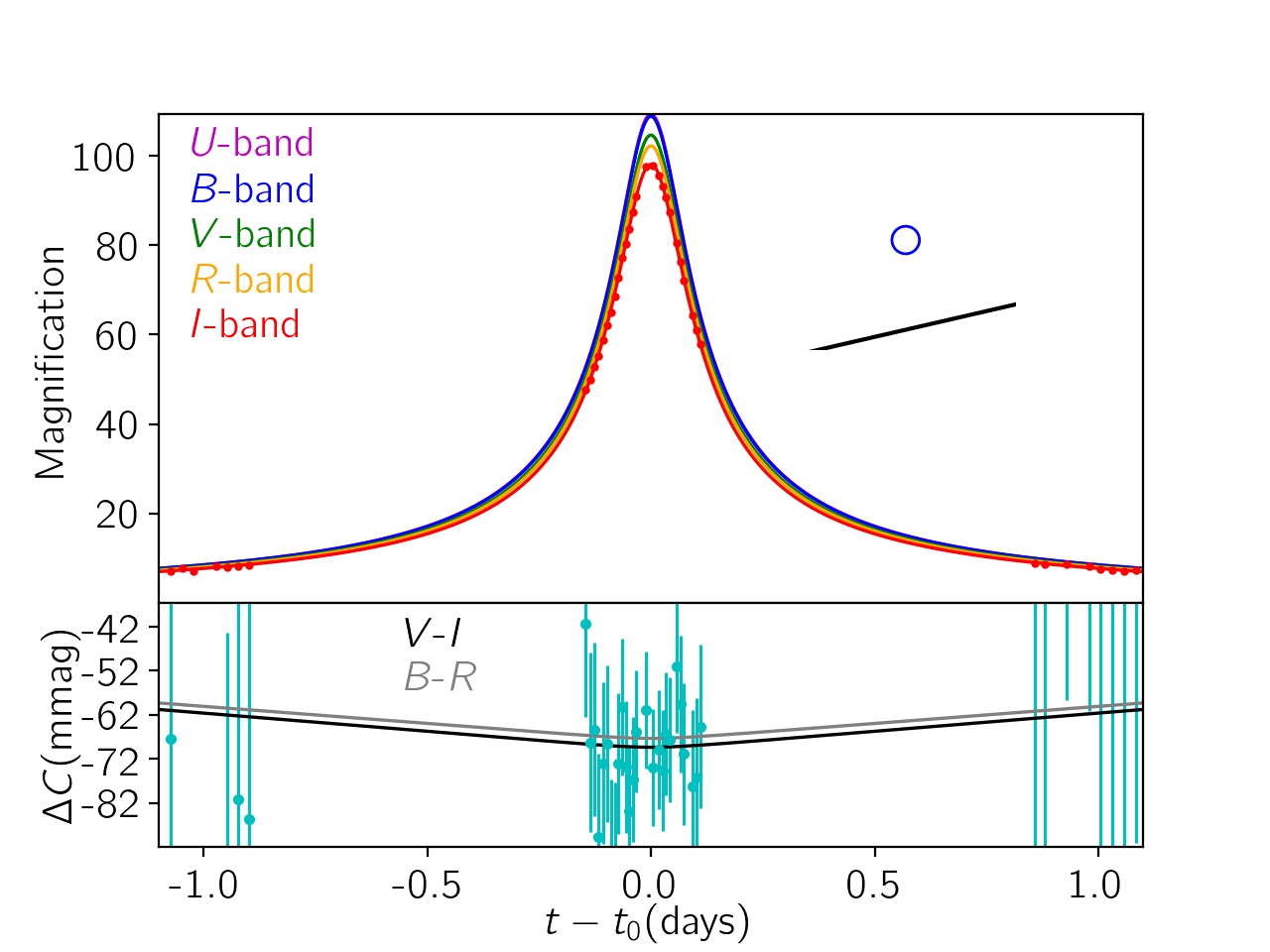}\label{fig0b}}
\subfigure[]{\includegraphics[width=0.48\textwidth]{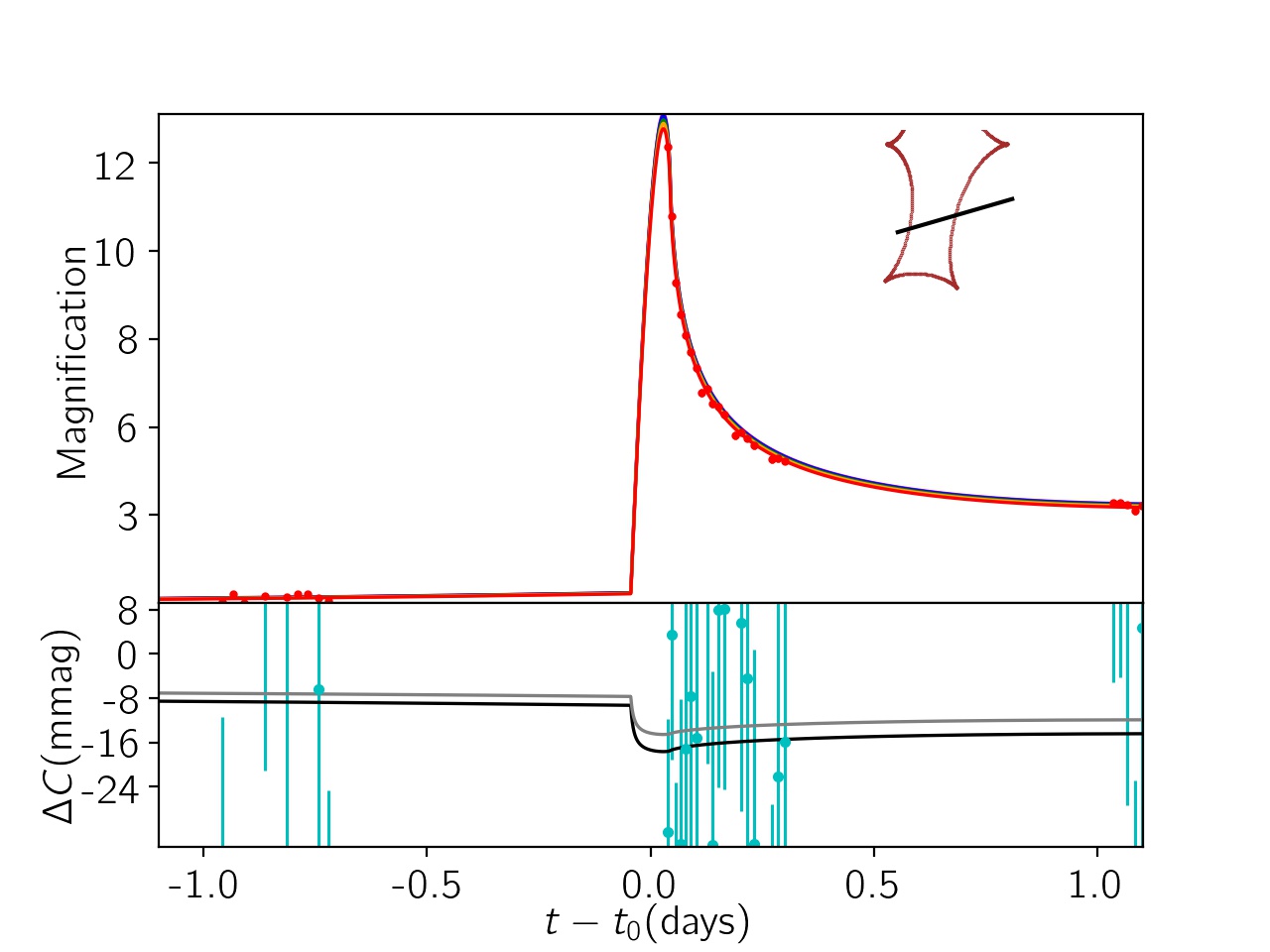}\label{fig0c}}
\subfigure[]{\includegraphics[width=0.48\textwidth]{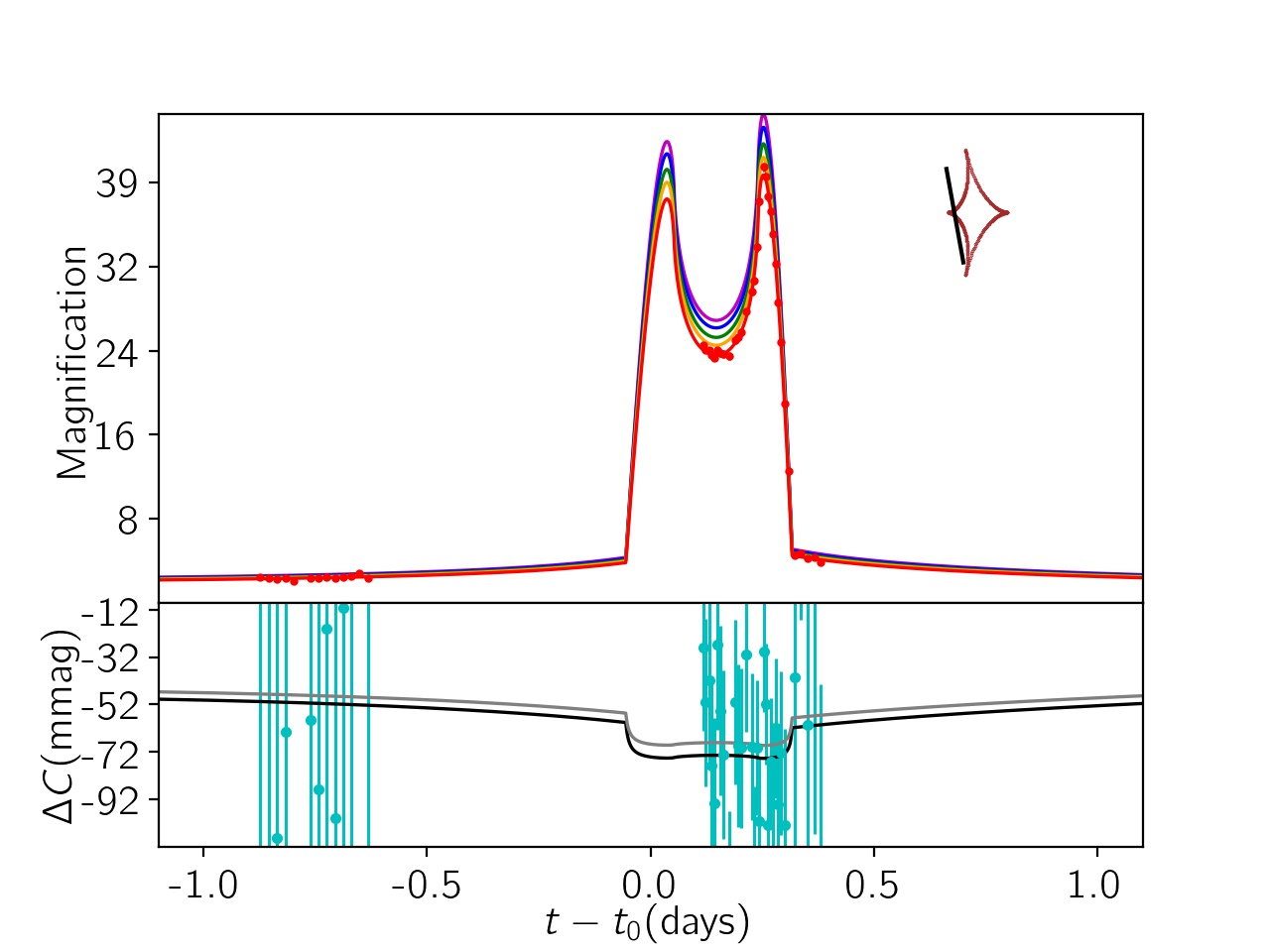}\label{fig0d}}
\caption{HM and CC microlensing lightcurves of blended source stars in the passband filters $UBVRI$, are shown by magenta, blue, green, orange, red colors, respectively. The perturbations in stellar colors $V$-$I$ and $B$-$R$ during the lensing effect are shown in the residual parts by solid (black and gray) curves, respectively. The source star parameters are listed in Table \ref{tabblend}. The synthetic data points by the LI camera at the Danish telescope in the $I$-band are shown with red filled circles. The corresponding data points over the residual parts (related to the $V$-$I$ color) are plotted by cyan filled points. In the two top panels, each right-hand inset shows the source star (blue circle) and the lens trajectory (black line) both projected on the lens plane. In the bottom panels, each right-hand inset shows the caustic curve (brown curve) and the source trajectory (black line) projected onto the lens plane.}\label{blendl}
\end{figure*}
\begin{table*}             
	\centering          
	\caption{The parameters of the lightcurves shown in Figure \ref{blendl}.}
	\begin{tabular}{c c c c c c c c c c c}
		\toprule[1.5pt]
		$~$&$\log_{10}[u_{0}]$&$t_{\rm{E}}$&$\log_{10}[\rho_{\ast}]$&$m_{\rm{base},~I}$&$T_{\rm{eff}}$& $b_{I}$ &$q$&$d$&$\xi$&$\Delta \chi^{2}$\\
		&&$\rm{(days)}$&&$\rm{(mag)}$&$\rm{(K)}$&&&&$\rm{(deg)}$&\\ 
		\toprule[1.5pt]
		\ref{fig0a}&  -2.39 & 5.5 &  -2.58 &  20.4 &5009 & 0.73 & - & - & 311 & 28\\%% corrected 25 Aug
		\ref{fig0b}& -2.08 & 9.8 &  -2.88 &  20.2 & 4653 & 0.81 & - & - & 193 & 547\\  %% corrected 25 Aug
		\ref{fig0c}& -0.43 & 10.1 & -2.43 & 18.9 & 5895 & 0.97 & 0.59 & 0.94 & -163 & 10\\%% corrected 25th Aug
		\ref{fig0d}& -0.95 & 7.8 &  -2.36 &  20.4 & 5132 & 0.84 & 0.61 & 0.68 & 100 & 132\\%% 25 th Aug
		\hline
	\end{tabular} 
	\tablefoot{Here, $d$ is the distance between two lenses normalized to the Einstein radius, $q$ is the mass ratio of binary lenses, $\xi$ is the angle between the horizontal axis and the source trajectory and $m_{\rm{base},~I}$ is the apparent magnitude of the source star in $I$-band at the baseline.} \label{tabblend}      
\end{table*}

\noindent Accordingly, the observing magnification factor by considering the blending effect is evaluated by: 
\begin{eqnarray}
A_{\rm b,~F}(t)=\frac{\mathcal{F}_{\ast,~F}~A(t)~+~\mathcal{F}_{\rm{b},~F} }{\mathcal{F}_{\ast,~F}~+~\mathcal{F}_{\rm{b},~F}}= b_{F}~A(t) + 1- b_{F}, \label{amags}
\end{eqnarray}
where, $b_{F}=\mathcal{F}_{\ast,~F}/(\mathcal{F}_{\ast,~F}+ \mathcal{F}_{\rm b,~F})$ is called the blending parameter and $\mathcal{F}_{\rm{b},~F}= \sum_{i=1}^{N_{\rm b}} \mathcal{F}_{i,~F}$ is the total flux due to all blending stars. $A(t)$ is the magnification factor of the source star. For point-like source stars in single microlensing events, the magnification factor is evaluated by Equation \ref{simpleA}. By considering finite-size of the source star, this magnification factor deviates from its simple model which will be studied in the next sections.

The blending parameter is unity if there is no blending stars. It also depends on the adopted filter. Hence, during a microlensing event the blending effect makes chromatic perturbations. We measure the variation in the stellar color during a microlensing event by:  
\begin{eqnarray}
\Delta C_{FF'} =C(t)-C_{\rm{base}}= -2.5 \log_{10}\left[\frac{A_{\rm b, ~F}}{A_{\rm b, ~F'}}\right],\label{deltac}
\end{eqnarray}
\noindent where $\Delta C_{FF'}$ is the variation in the stellar color during a microlensing event with respect to the baseline, $F,~F' \in UBVRI$, and $C= m_{F}-m_{F'}$ is the apparent stellar color, i.e., the difference of the stellar apparent magnitude in filters $F,~F'$. Here, we have $m_{F}=m_{\rm{base},~F} -2.5 \log_{10}\left[A_{\rm b,~F}\right]$ and $C_{\rm{base}}=m_{\rm{base},~F}-m_{\rm{base},~F'}$.

In order to evaluate the above chromatic perturbations due to blending effect, we simulate large number of microlensing events toward the Galactic bulge. Because, most of nowadays microlensing observations are done in these directions, e.g., toward the Baade's window \citep[e.g.,][]{1999Paczynski}. However, in the real observations highly blended microlensing events are eliminated \citep[see, e.g.,][]{2017NaturMroz, 2019ApJSMroz}, because the high blending effect makes a sever degeneracy while modeling microlensing events \citep[see, e.g., ][]{2007Smith, 2020AJMroz, 2021sajadian}. Hence, in the simulation we exclude the high-blended events with $b_{I}<0.2$. By simulating microlensing events toward the Baade's window with the coordinate $(l_{\rm G},~b_{\rm G})=(1^{\circ},~-4^{\circ})$, the average blending parameters in different filters are estimated $\left<b_{U}\right>,~\left<b_{B}\right>,~\left<b_{V}\right>,~\left<b_{R}\right>,~\left<b_{I}\right>\simeq 0.71,~0.76,~0.75,~0.73,~0.72$, respectively. The blending parameter in fact depends on observing directions.

\noindent Considering these blending parameters, variations in the stellar colors $V$-$I$ and $B$-$R$ are plotted in Figure \ref{blendd} versus the magnification factor. According to this plot, we list some key points here. 

\begin{itemize}
\item The blending-induced chromatic variations increase by enhancing the magnification factor. For the magnification factor higher than $\gtrsim 40$, their increasing slopes decrease and tend to zero. However, the maximum chromatic perturbations happen at the magnification peak. 
	
\item The time scale of the chromatic perturbations due to blending effect is of order $t_{\rm E}$ (the Einstein crossing time which determines the time scale of a microlensing event). Because, by enhancing the magnification factor from unity the chromatic perturbations increase as well.
	
\item For transit microlensing events, the magnification factor (by considering the finite-source effect) remains almost constant while the lens object is passing over the source disk, which results constant chromatic perturbations.
\end{itemize}

In Figure \ref{blendl}, four examples of HM and CC microlensing lightcurves by considering the blending effect are represented. In these plots, the magnification factors in the filters $UBVRI$ are plotted with magenta, blue, green, orange, red colors, respectively. In order to calculate the magnification factor in binary microlensing events, we use the $\rm{RT}$-model \footnote{\url{http://www.fisica.unisa.it/gravitationastrophysics/RTModel.htm}} which was developed by V.~Bozza \citep{Bozza2018, Bozza2010, Skowron2012}.
	
\noindent In these lightcurves, the time of closest approach defines time zero. The residual parts show the variation in the stellar colors $V$-$I$ (black solid curves) and $B$-$R$ (gray solid curves) in unit of magnitude (Equation \ref{deltac}). In the top panels, the right-hand insets show the source disk (blue circles) and the lens trajectory (black lines) both projected on the lens plane. In the bottom panels (binary CC microlensing events), the right-hand insets represent the caustic (brown curve) and the source trajectory projected onto the lens plane (black line). As mentioned above, these chromatic perturbations around the magnification peak vary very slowly.

\begin{table*}
\caption{The LI efficiencies for detecting chromatic perturbations in HM single and CC binary microlensing events and the properties of these events.}             
\centering          
\begin{tabular}{ccccccccccc}    
\toprule[1.5pt]
$~$&$\epsilon_{\rm l}$&$\epsilon_{\rm h}$&$\left<\Delta C_{VI}\right>$&$\left< \Delta C_{BR}\right>$ & $\log_{10}\left[\left<u_{0}\right>\right]$ &  $\left<b_{I}\right>$& $f_{100}$ & $f_{200}$& $f_{500}$& $\rm{No.}$\\
&$[\%]$&$[\%]$&$\rm{(mmag)}$&$\rm{(mmag)}$& & & $[\%]$ & $[\%]$& $[\%]$ &\\  
\toprule[1.5pt]
\multicolumn{11}{c}{$\rm{HM}~\rm{microlensing}$}\\
$\rm{BL}$ & 61.1 & 48.7  & 94.2 & 105.0  & -2.31 & 0.75 & 76.5& 38.2 & 10.8 & 48955\\
$\rm{BL}~\&~\rm{LD}$ & 62.2 & 50.0  & 98.2 & 112.5  & -2.31 & 0.75 & 75.4 & 37.6 & 11.1 & 58329 \\
$\rm{BL}~\&~\rm{LD}~\&~\rm{CGP}$ & 64.2 & 53.1  & 58.7 & 79.1  & -2.31 & 0.81 & 75.2 & 35.9 & 7.7 & 30932\\
$\rm{BL}~\&~\rm{LD}~\&~\rm{SS}$ & 62.7 & 51.0  & 97.6 & 111.0  & -2.31 & 0.75 & 75.2 & 37.7 & 11.2 & 1721\\
$\rm{BL}~\&~\rm{GD}$ &  62.2 & 53.7  & 35.7 & 35.9  & -2.30 & 0.94 & 93.2 & 43.6 & 8.6 & 3644 \\
\hline 
\multicolumn{11}{c}{$\rm{CC}~\rm{microlensing}$}\\
$\rm{BL}$ & 26.1 & 17.3  & 97.9 & 111.1  & -0.75 & 0.74 & 19.1& 3.5 & 0.1 & 6350\\
$\rm{BL}~\&~\rm{LD}$  &  30.8 & 21.7  & 106.6 & 134.2  & -0.75 &  0.77 & 18.7 & 3.1 & $<$0.07 & 1333 \\
$\rm{BL}~\&~\rm{LD}~\&~\rm{CGP}$& 33.2 & 22.7  & 108.8 & 138.1  & -0.71 & 0.78 & 17.9 & 2.9 & $<$0.1 & 712\\
$\rm{BL}~\&~\rm{LD}~\&~\rm{SS}$ & 35.8 & 26.8  & 104.9 & 125.6  & -0.72 & 0.80 & 14.7 & 1.1 & $<$0.2 & 584 \\
$\rm{BL}~\&~\rm{GD}$ &  48.2 & 40.9  & 51.7 & 53.9  & -0.67 & 0.93 & 11.3 & 0.8 & 0.8 & 433 \\
\hline
\end{tabular} 	
\tablefoot{$\rm{BL},~\rm{LD},~\rm{CGP},~\rm{SS}$ and $\rm{GD}$ refer to the blending, limb-darkening, close-in giant planet companion, stellar spot and gravity-darkening effects, respectively. $\left<\Delta C_{VI}\right>$ and $\left< \Delta C_{BR}\right>$ are the average values of maximum chromatic deviations in detectable (simulated) microlensing events. $f_{100}$, $f_{200}$ and $f_{500}$ represent the fraction of the simulated events with detectable chromatic perturbations, and with the maximum magnification factor higher than $100$, $200$ and $500$, respectively. The last column $(\rm{No.})$ reports the number of simulated events with detectable perturbations.}\label{tab_result}      
\end{table*}
\noindent In order to study the detectability of the resulted chromatic perturbations, we generate synthetic data points corresponding to observations with the lucky imaging (LI) camera at the Danish $1.54$ m telescope at ESO's La Silla observatory in Chile. In Figure \ref{blendl}, the data points in the $I$-band are shown as red points on the magnification lightcurves (top parts) and the resulting perturbations in the stellar color $V$-$I$ (given by Eq. \ref{deltac} and in magnitude units) are plotted as cyan points on the residual parts. 

\noindent Some scatters were added to the data points given in filter $F$ by using a Gaussian distribution with a width equal to the error bars, $\delta_{F}$. In Appendix \ref{append}, we explain how to determine the error bars of the synthetic data points taken with the LI camera. Subtracting the two magnitudes, in the $F,~F'$-bands, from each other results the error bar in the stellar color as $\delta_{C}=\sqrt{\delta_{F}^{2}+\delta_{F'}^{2}}$. The cadence between data points from on-going microlensing events, taken by this camera, depends on the magnification factor and can change in the range $\tau \in [2~\rm{min},~2~\rm{hours}]$, as explained  in \citet{Dominik2010, Sajadian2016A}.

\noindent The parameters of these lightcurves are given in Table \ref{tabblend}. In this table, at its last column the value of $\Delta \chi^{2}$ is given which is $\Delta \chi^{2}= \Big|\chi^{2}_{\rm{real}}-\chi^{2}_{\rm{base}}\Big|$, i.e., the difference between the $\chi^{2}$ values from fitting the real models and the baseline amount for the source color versus time.   
	
Comparing Figures \ref{fig0a} and \ref{fig0b}, the higher blending effects make larger chromatic deviations. However, the blending effect decreases maximum magnification factors which restricts the achievable single-to-noise ratio. Hence, detecting these chromatic perturbations in highly blended microlensing events is hard, unless the source star at the baseline is very bright.
	
In order to evaluate the detectability and properties of the chromatic perturbations in HM microlensing events with the LI camera, we have generated a big ensemble of these (HM and CC) events. For simulating HM events, we choose the lens impact parameter uniformly from the range $[0,~0.01]$ (which constitute around one per cent of all microlensing events). We simulate their lightcurves during a two-day time interval around their peaks. The start time of simulating data points is chosen uniformly from the range $[0,~24]$ hours, and we assume that the observations can be done during $6.5$ hours in a day. For simulating binary microlensing events, we only simulate the CC features and generate the synthetic data points in a two-day time interval around them.  
	
We consider $\Delta \chi^{2}>100,~200$ as detectability criterion for the chromatic perturbations with low and high sensitivities in the simulated lightcurves. The results from this Monte-Carlo simulation and other simulations, which will be explained in next sections, are summarized in Table \ref{tab_result}. In this table, $\epsilon_{\rm h}$ and $\epsilon_{\rm l}$ are the LI efficiencies for detecting chromatic perturbations with high and low sensitivities. These chromatic perturbations are potentially generated due to different physical effects or their combination (identified with the symbol $\&$), as specified in different rows of the table. Two next columns of this table also represent average values of maximum chromatic deviations in detectable (simulated) microlensing events. The next column reports the average of the lens impact parameter of detectable microlensing events. $f_{100}$,~$f_{200}$, and $f_{500}$ represent the fraction of the simulated events with detectable chromatic perturbations, and with the maximum magnification factor higher than $100$, $200$ and $500$, respectively. The last column, $\rm{No.}$, declares the number of simulated events with detectable perturbations.
\begin{figure*}
\centering
\includegraphics[width=0.49\textwidth]{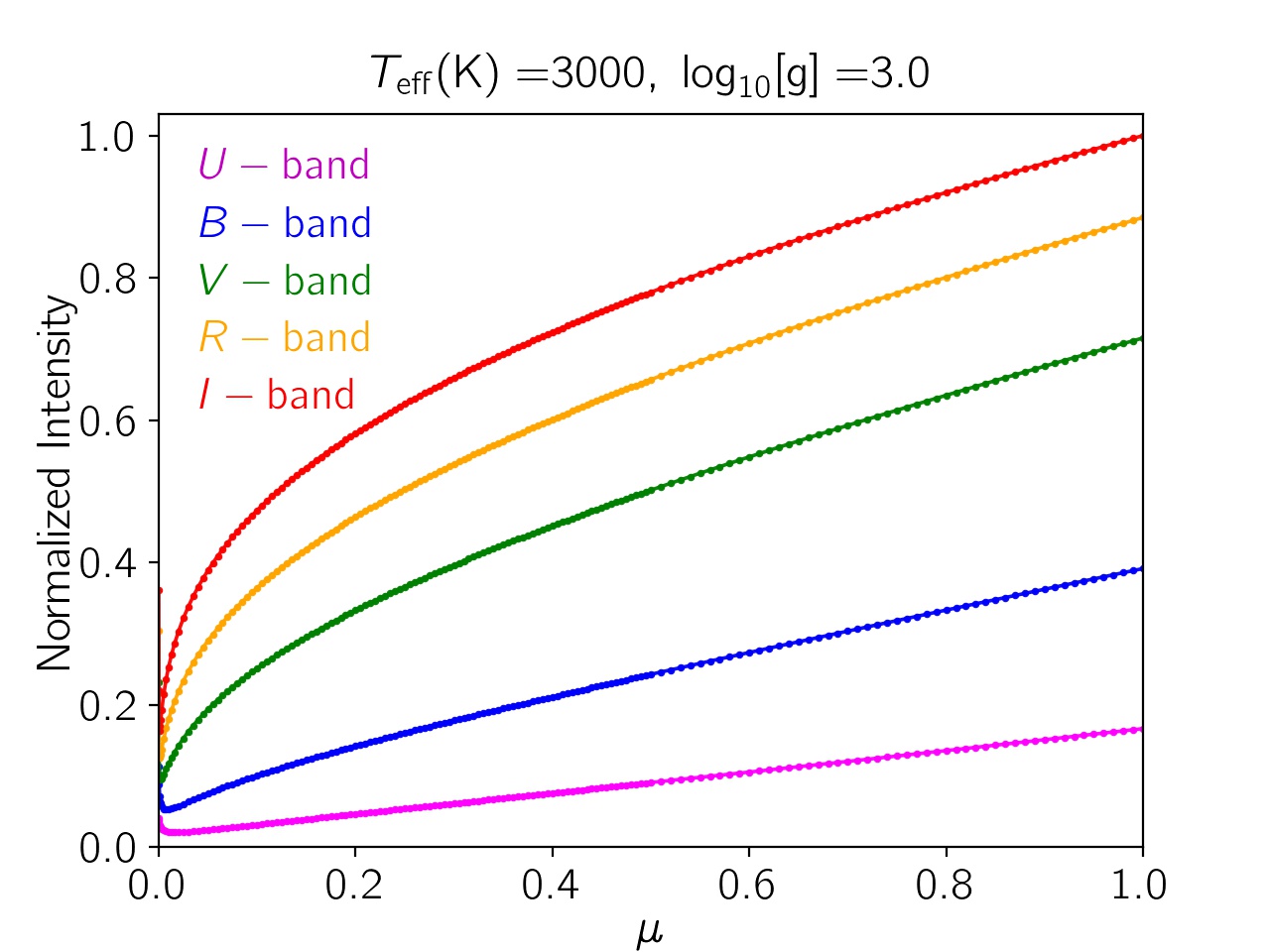}
\includegraphics[width=0.49\textwidth]{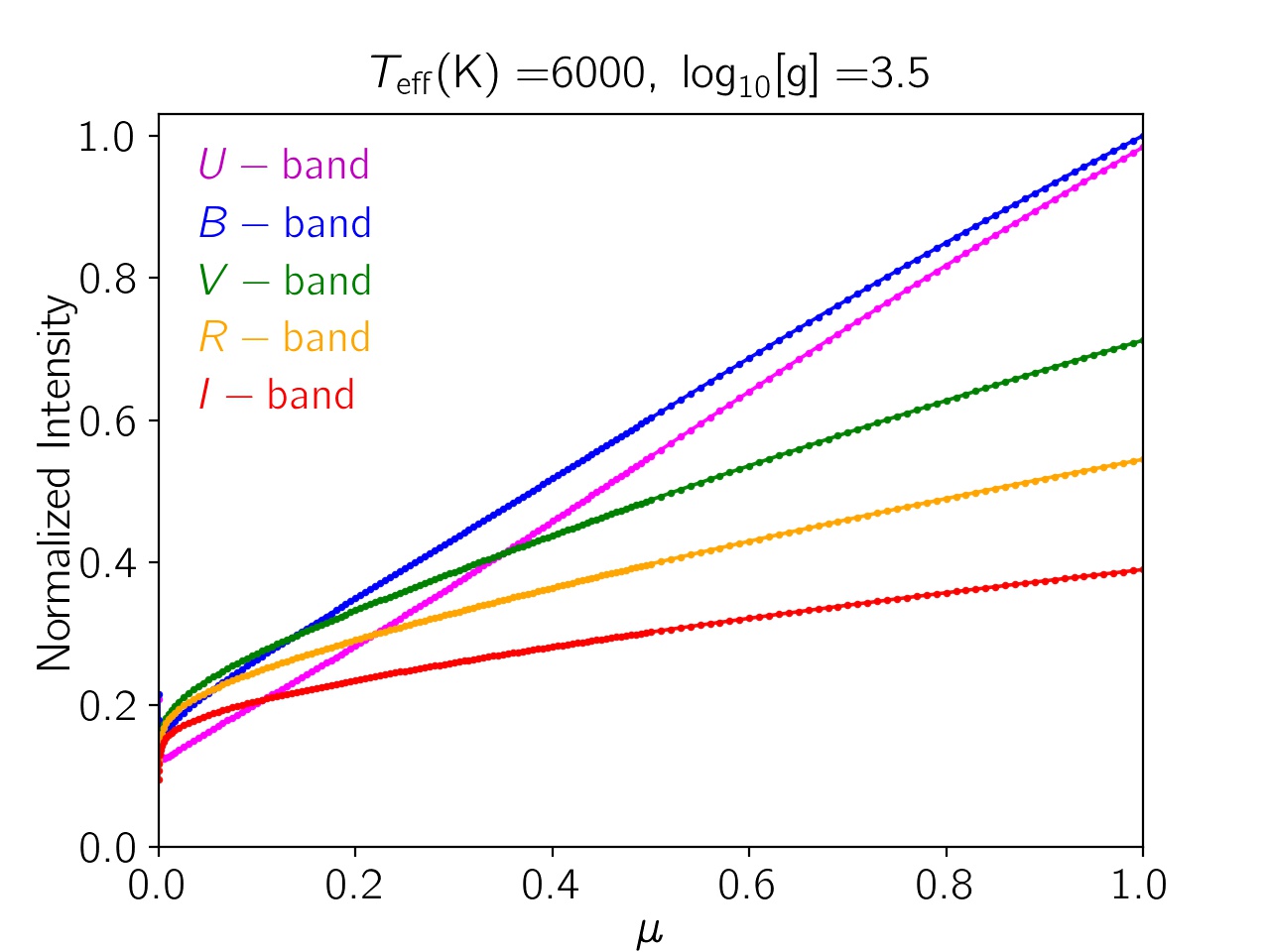}
\includegraphics[width=0.49\textwidth]{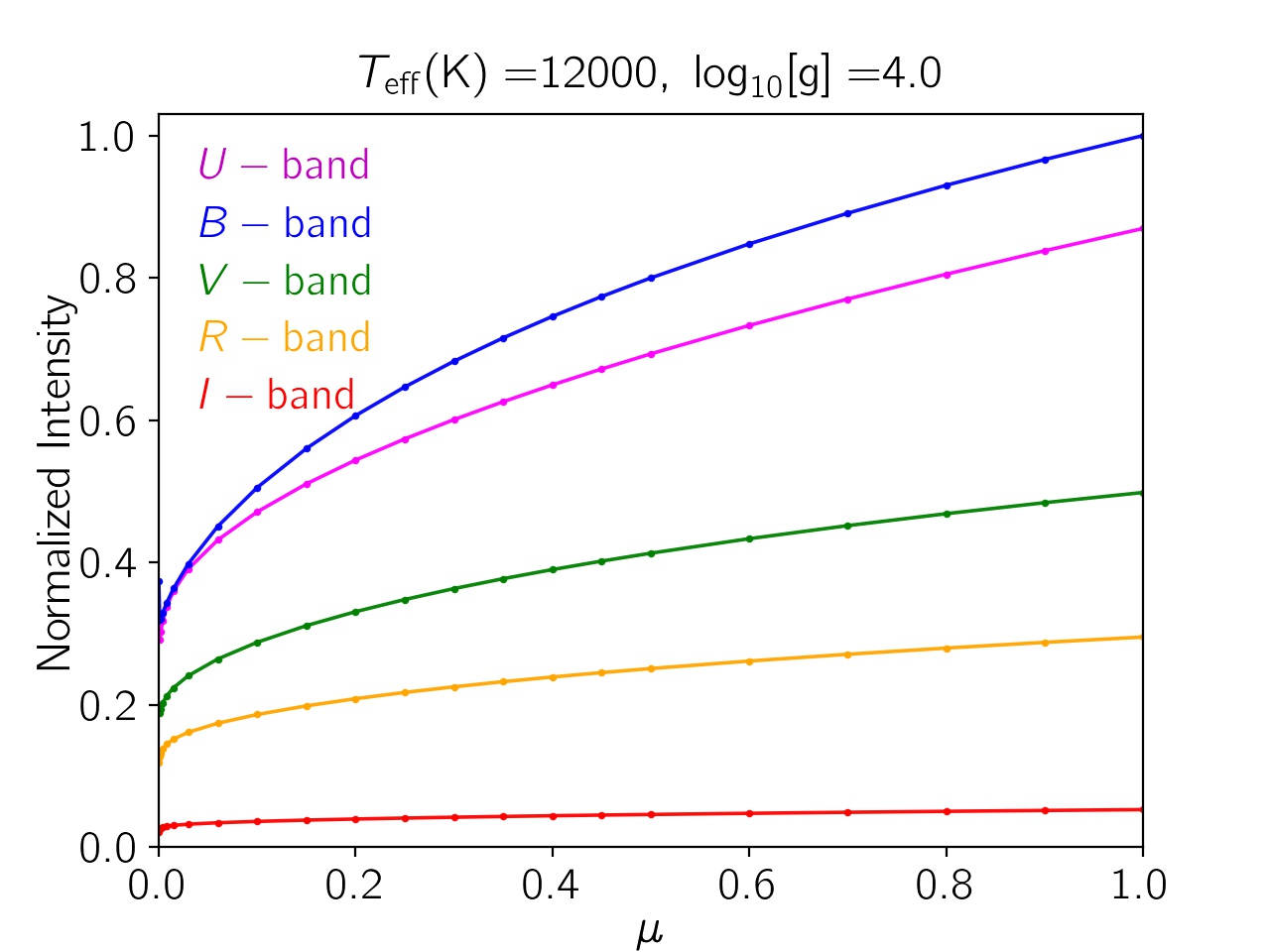}
\includegraphics[width=0.49\textwidth]{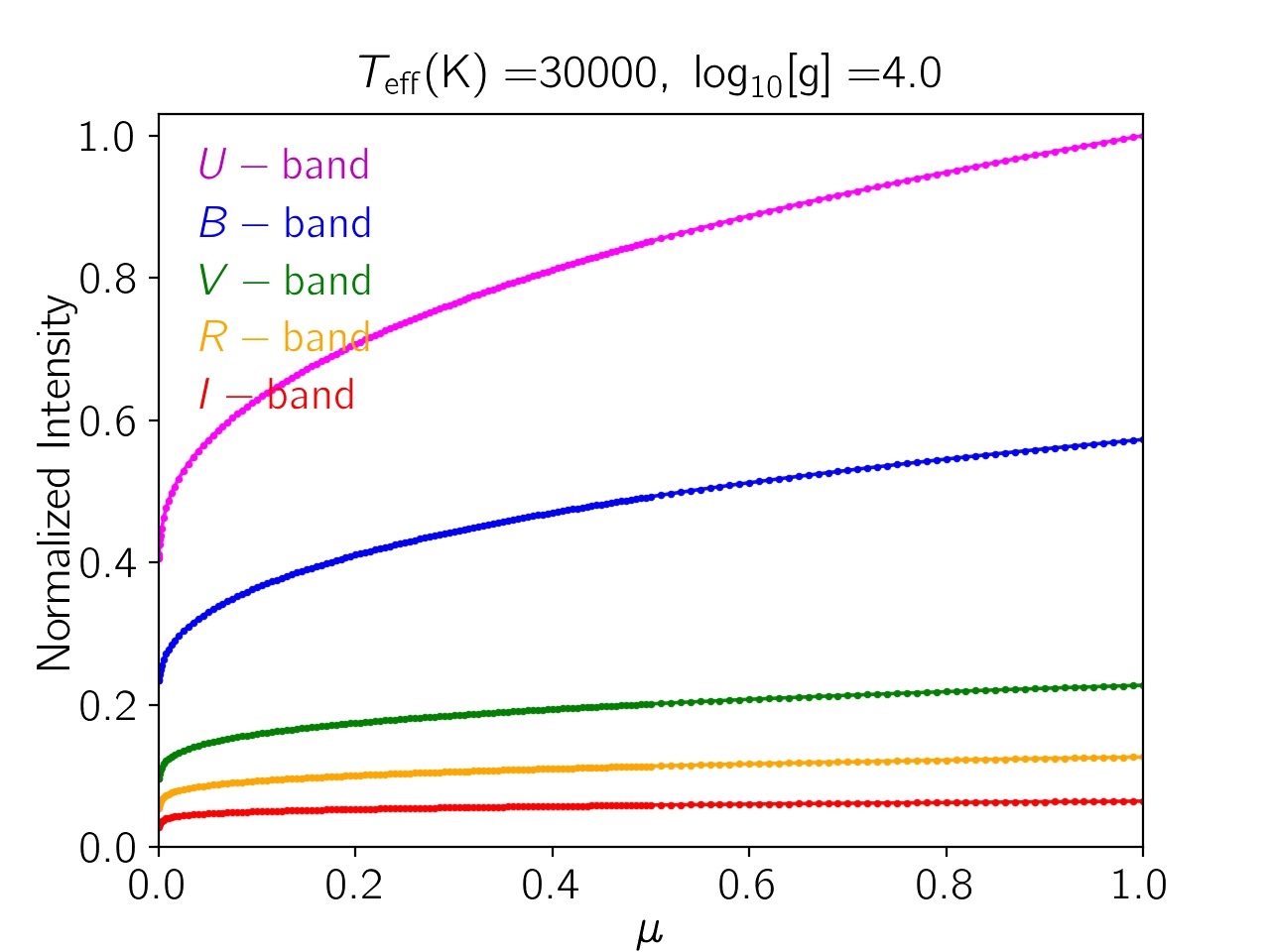}
\caption{The emergent radiation intensities for four stars with different temperature and the surface gravity values, versus $\mu$, in the standard filters $UBVRI$ as shown with magenta, blue, green, orange and red curves.}\label{one}
\end{figure*}
\begin{figure*}
	\centering
	\includegraphics[width=0.49\textwidth]{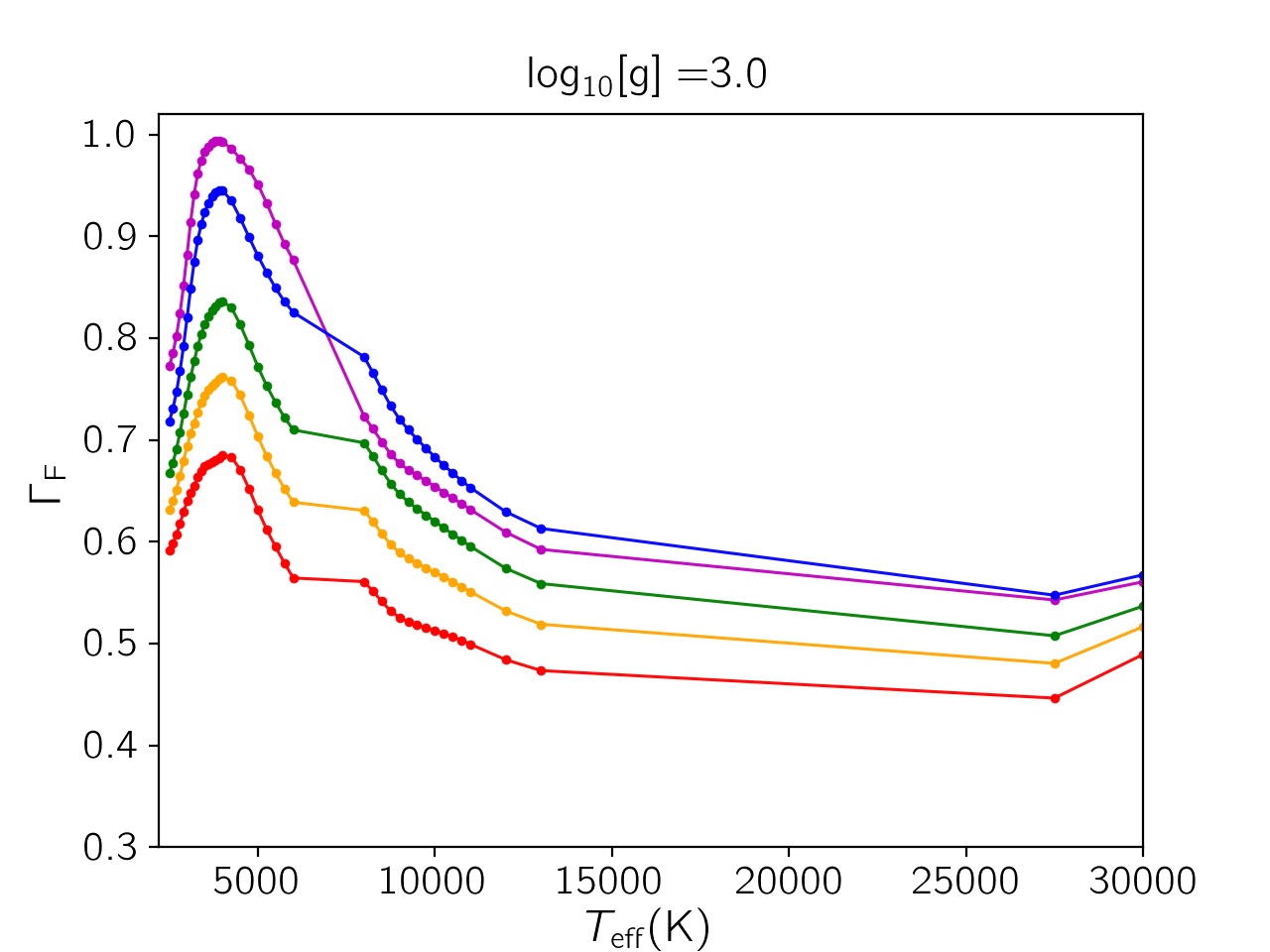}
	\includegraphics[width=0.49\textwidth]{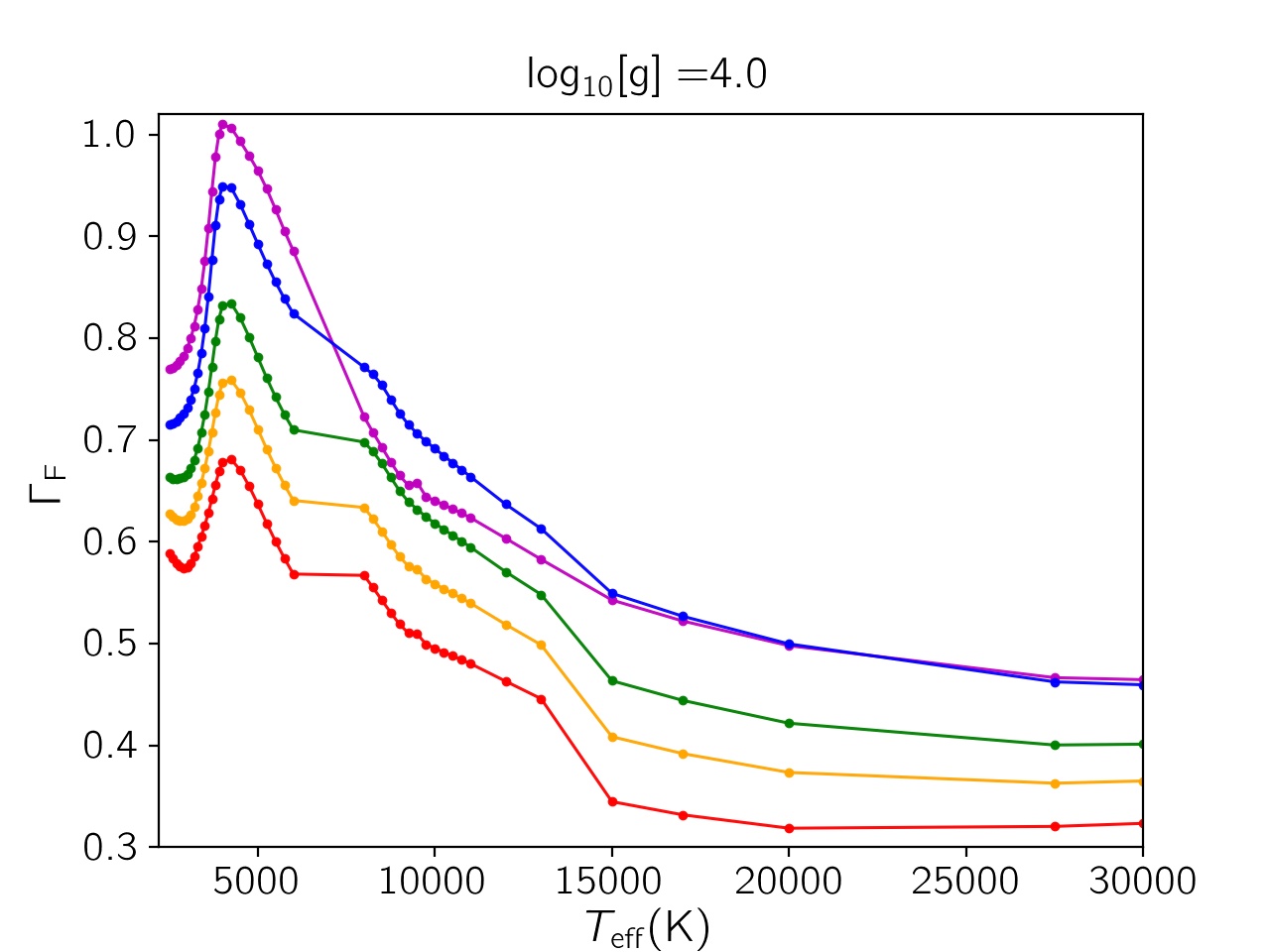}
	\caption{The LD coefficients, $\Gamma_{F}$, in different filters $F \in UBVRI$ (shown with magenta, blue, green, orange and red colors, respectively) for stars with different effective surface temperatures and two values of the surface gravity (given at the top of the plots).}\label{two}
\end{figure*}

Accordingly, the chromatic deviations due to the blending effect can be detected rather in HM microlensing events (with the maximum magnification higher than $100$) than CC binary events. In the next section, we add another source for chromatic perturbations which is the LD effect.

%%%%%%%%%%%%%%%%%%%%%%%%%%%%%%%%%%%%%%%%%%%%%%%%%%%%%
\section{Limb-darkening effect}\label{limb}
The gradual decrease in brightness of the stellar disks from center to limb, as seen by the observer, is called limb-darkening. The line of sight through the atmosphere is larger on the limb than at the center of the disk, and hence the light from the limb is dominated by higher layers in the atmosphere than the light reaching the observer from the center of the disk. Since the higher layers are generally cooler than the deeper layers in the atmosphere, the light from the limb is generally more redder than the light from the center. In other words, photons can escape the stellar photosphere when the optical thickness is about one for a given wavelength, so photons observed in the center of a stellar disk comes from a deeper layer of the stellar atmosphere than photons observed toward the limbs.

\noindent The detailed calculation of this dependency needs modeling of the stellar atmospheres and numerically solving the transfer equations \citep{Code1950,chandrasekhar1960}. Such simulations have been done by P.~Harrington in 2017 \footnote{\url{https://www.astro.umd.edu/~jph/Stellar_Polarization.html}} \citep{Harrington1970,Harrington2015,Harrington2017}. He considered several models of the stellar atmospheres, such as the \textit{MARCS} models for cool stars with an effective temperature range $[2500,~8000]$ K, and $\log[g]= 3,~3.5,~4,~4.5,~5$ \citep{MARCS2008,MARCS2017}, the \textit{TLUSTY} models for hot stars \citep{TLUSTY2003,TLUSTY2007} and the \textit{STERNE} and \textit{KURUCZ} models for stars within the temperature range $[8000,~15000]$ K \citep{kurucz1993,Kurucz2003}. Accordingly, he offered the polarized and total emergent continuum radiation at different wavelengths versus $\mu=\sqrt{1-\rho ^{2}}$, where $\rho=R/R_{\ast}$, $R$ is the radial distance over the source disk from its center and $R_{\ast}$ is the stellar radius. 

\noindent Using his output, we integrate the emergent radiation, $B(\mu,~\lambda)$, over the wavelength by considering the transmission function for the standard filters $F \in UBVRI$, as follows:
\begin{eqnarray}
I_{F}(\mu)=\int_{0}^{\infty} d\lambda~K_{F}(\lambda-\lambda_{0})~B(\mu, \lambda), 
\end{eqnarray}
\noindent where $K_{F}(\lambda-\lambda_{0})$ is the transmission function for the filter under consideration. For four stars with effective temperatures of $3000,~6000,~12000,$~and~$30000$ K, we plot their normalized intensities in the different filters $UBVRI$ with magenta, blue, green, orange and red colors versus $\mu$ in Figure \ref{one}. These plots show the dependence of the stellar radiation on the used filters. All of these curves have a similar trend, with their largest value at the source center and decreasing slowly from center to limb. For hotter stars the higher emergent intensities occur at shorter wavelengths. In order to study for what kind of stars the contrast of stellar radiation in different filters is highest, we fit linear functions to the stellar radiation and compare their coefficients.

Generally, for the emergent radiation versus $\mu$, a very simple and known function is considered, which is:  
\begin{eqnarray}\label{linear}
I_{F}(\mu)=I_{0,~F} \Big[1- \Gamma_{F} (1- \mu) \Big],
\end{eqnarray}
\noindent where $\Gamma_{F}$ is the so-called LD coefficient which depends on the adapted filter. We estimate $\Gamma_{F}$ for the standard filters $UBVRI$ for the different types of stars with different effective temperature and surface gravity and plot the results in Figure \ref{two}. The LD coefficients in the $UBVRI$ filters are shown by magenta, blue, green, orange and red colors, respectively. 

The emergent radiation curves do not behave linearly, especially for $\mu \lesssim 0.1$ (see Fig. \ref{one}). Hence, the linear function given by Equation \ref{linear} mostly describes the emergent radiation for $ \mu \gtrsim 0.1$. The largest contrast between the LD coefficients in different filters happens for stars with intermediate temperatures around $T\sim 4000$-$5000$ K. Changes in surface gravity does not have a significant impact on the emergent radiation contracts between different filters, due to the corresponding low sensitivity of the absorption coefficients on gravity.

Dependence of the emergent radiation in the different filters under consideration causes the stellar color to change in microlensing events, while the lens is passing over (or close to) the disk of the source star. In order to study the variation in the stellar color, we inset the lensing magnification factor in the stellar flux integration over the source disk as:  
\begin{eqnarray}
\mathcal{F}_{F}(t)=\frac{1}{4\pi D_{\rm s}^{2}} \int_{0}^{\infty} d\lambda K_{F}(\lambda-\lambda_{0})\int dx~dy~B(\mu, \lambda) A(u_{x}, u_{y}, t), 
\end{eqnarray}
\noindent where, $(x,~y)$ is the location of each element over the surface of the source star, so that $R=\sqrt{x^{2}+y^{2}}$, $(u_{x},~u_{y})$ is the location of that element with respect to the lens position projected on the lens plane and normalized to the Einstein radius. If there is no blending effect, the magnification factor of a limb-darkened source star in the adopted filter $F$ is given by:  
\begin{eqnarray}
A_{\rm{LD},~F}(t)= \frac{\mathcal{F}_{F}(t)}{\mathcal{F}_{\rm{base},~F}},
\end{eqnarray}
\noindent where, $\mathcal{F}_{\rm{base},~F}$ is the stellar flux measured in the filter $F$ at the baseline. We expect that the largest variation in the stellar colors due to LD effects occurs during HM microlensing events (especially when the lens is directly transiting the source disk) and in CC binary microlensing events. This is because, when the lens impact parameter is larger than the source radius, or the source is far from the caustic line, the finite-source effect is small and all elements over the source disk have almost the same magnification factor. By including the blending effect the magnification factor will be:  
\begin{eqnarray}
A_{\rm{b}, \rm{LD},~F}(t)= b_{F}~A_{\rm{LD},~F}(t)~+~1- b_{F}. \label{ALD}
\end{eqnarray}

\begin{figure*}
\centering
\subfigure[]{\includegraphics[width=0.48\textwidth]{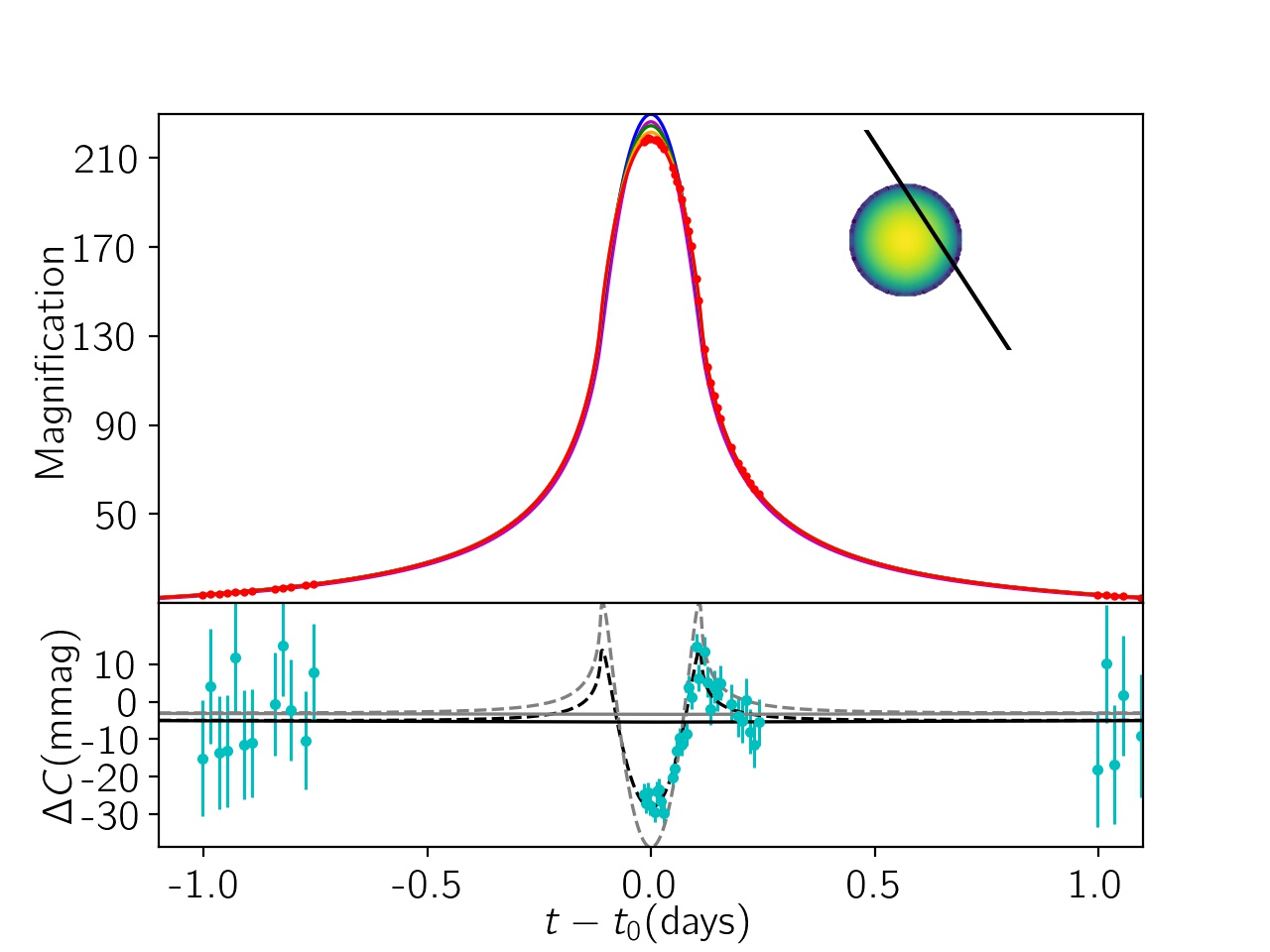}\label{fig2a}}
\subfigure[]{\includegraphics[width=0.48\textwidth]{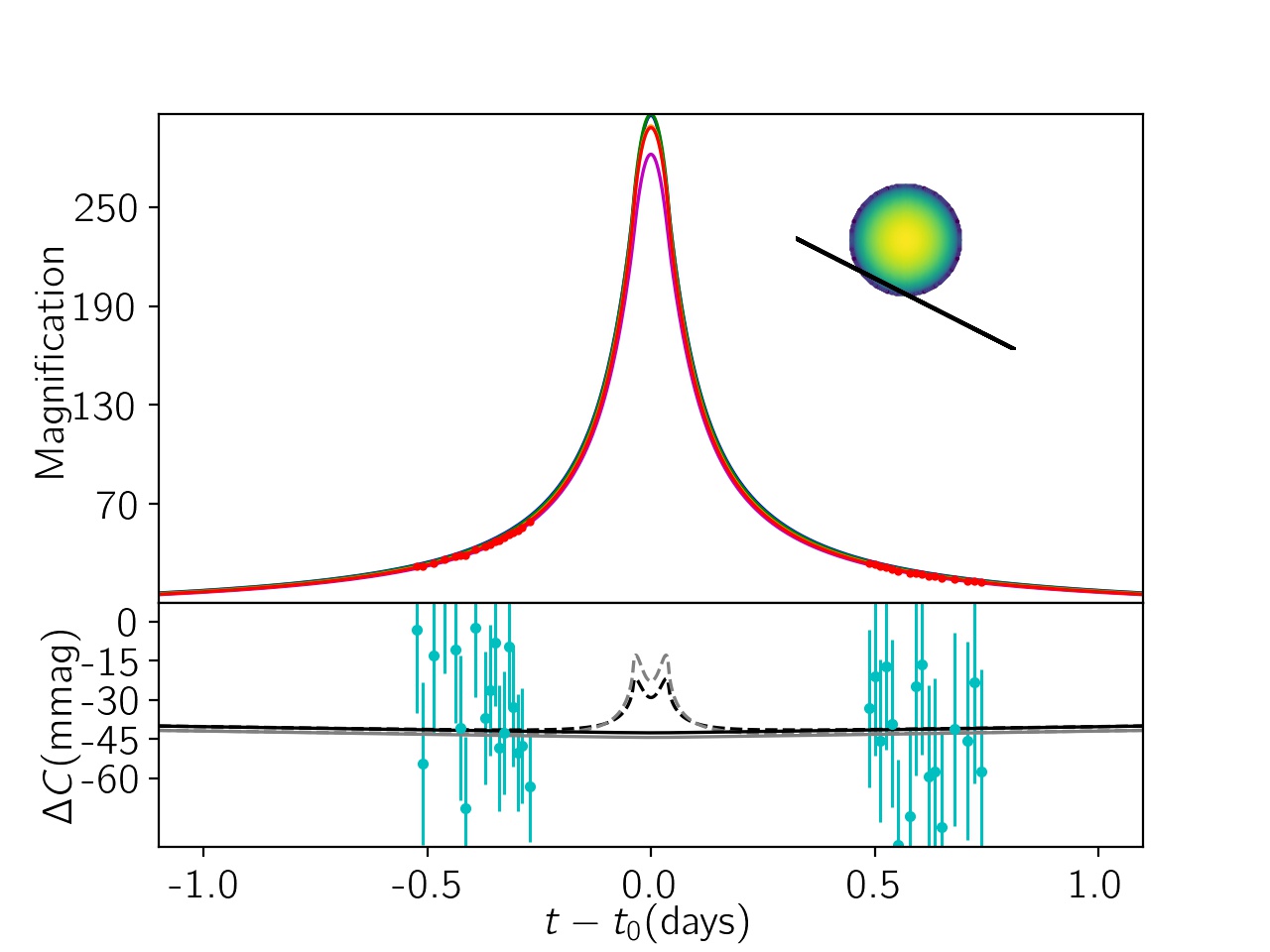}\label{fig2b}}
\subfigure[]{\includegraphics[width=0.48\textwidth]{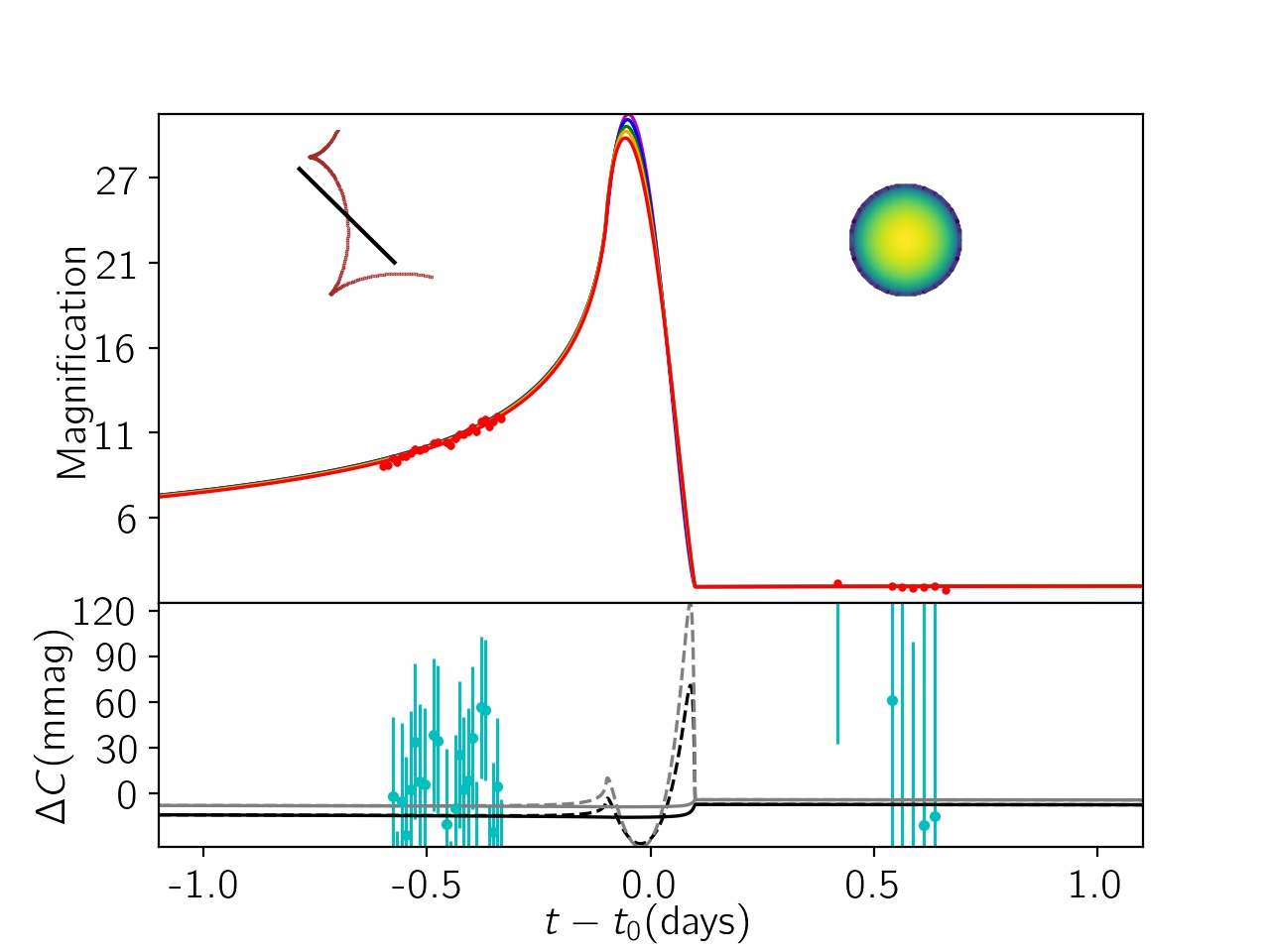}\label{fig2c}}
\subfigure[]{\includegraphics[width=0.48\textwidth]{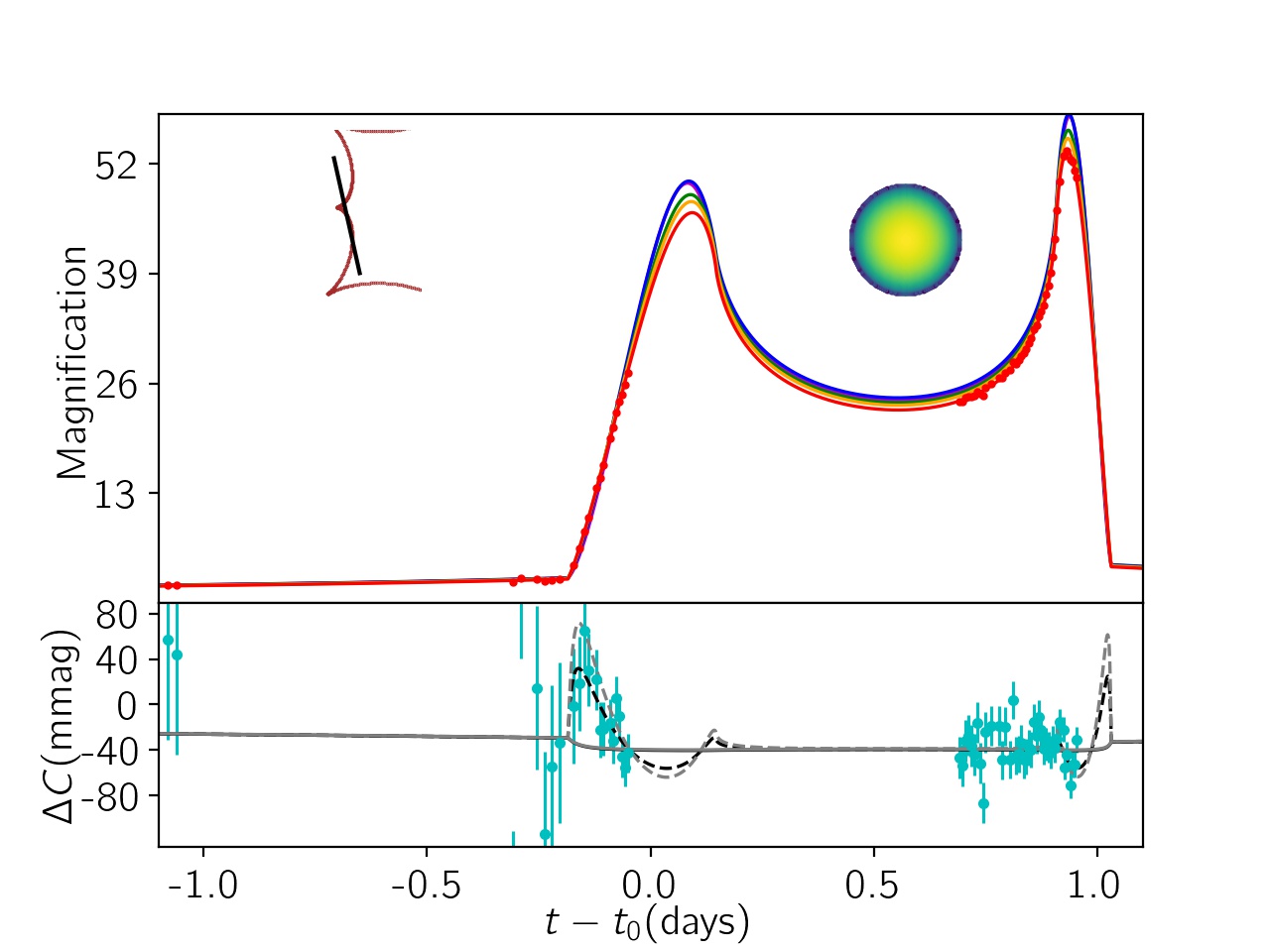}\label{fig2d}}
\caption{Similar to Figure \ref{blendl}, but for limb-darkened and blended source stars. The lensing parameters are listed in Table \ref{tab1}. In the residual parts, the perturbations in stellar colors $V$-$I$ and $B$-$R$ during the lensing effect due to LD and BL effects are shown by dashed (black and gray) curves, respectively. The corresponding chromatic BL-induced perturbations are shown by solid curves. The right-hand insets (colored schemes) represent brightness profiles over limb-darkened source disks.}\label{three}
\end{figure*}

Four examples of HM and CC microlensing lightcurves from limb-darkened and blended source stars are plotted in Figure \ref{three}. Some details about these plots can be found in the caption of Figure \ref{blendl}. In these lightcurves, the right-hand colored schemes represent the stellar surface brightness. In the residual parts, the perturbations in stellar colors $V$-$I$ (black) and $B$-$R$ (gray) during the lensing effect, due to LD and BL effects are shown by dashed curves, respectively. The corresponding chromatic perturbations due to only the BL effect are shown by solid curves. The parameters of the source stars responsible for the lightcurves are listed at Table \ref{tab1}.
\begin{table*}
	\caption{The parameters of the lightcurves shown in Figure \ref{three}.}             
	\label{tab1}      
	\centering          
	\begin{tabular}{ccccccccccc}
		\toprule[1.5pt]
		$~$&$\log_{10}[u_{0}]$&$t_{\rm{E}}$&$\log_{10}[\rho_{\ast}]$&$m_{\rm{base},~I}$&$T_{\rm{eff}}$ & $b_{I}$&$q$&$d$&$\xi$&$\Delta \chi^{2}$\\
		&&$\rm{(days)}$&&$\rm{(mag)}$&$\rm{(K)}$&&&&$\rm{(deg)}$&\\ 
		\toprule[1.5pt]
		\ref{fig2a}&  -2.36 & 14.6 &  -2.05 &  18.4 & 4849 & 0.99 & - & - & 303 &  1016\\%% 10
		\ref{fig2b}& -2.64 & 31.6 &  -2.59 &  20.4 & 4625 & 0.52& - & - & 153 & 55 \\%%% 27
		\ref{fig2c}& -0.51 & 38.3 &  -2.79 &  20.1 & 5453 & 0.98 & 0.76 & 1.13 & 135 &  0 \\%%% 3 corected
		\ref{fig2d}&  -0.78 & 17.1 &  -2.54 & 19.4 & 5729 & 0.85 & 0.30 & 1.16 & 102 &  310\\ %% 2_ corrected
		\hline
	\end{tabular}%\tablefoot{}
\end{table*}

We see from these plots that the peak and width of the microlensing lightcurves depend on the LD coefficients, and as a result also the filter under consideration. When the lens is passing very close to the source center $(u_{0}\simeq0)$, the value of the magnification peak is:
\begin{eqnarray}
A_{b,~LD,~F,~\rm{max}}(=A_{\rm m})\simeq \frac{2~b_{F}}{\rho_{\ast}} \frac{1-\Gamma_{F}(1-\pi/4)}{1-\Gamma_{F}/3},
\end{eqnarray} 
\noindent where, $\rho_{\ast}$ is the projected source radius normalized to the Einstein radius. The amount of the variation of the magnification peak in different filters is scaled by $\rho_{\ast}^{-1}$, i.e., the smaller the projected radii of the source stars, the larger the variations in the magnification factor in the different filters. This magnification peak is an increasing function of $\Gamma_{F}$. We see from Figure \ref{two} that for a given source star, the coefficient of $\Gamma_{F}$ is smaller at longer wavelengths, which results in more flattened lightcurves. However, if the lens is crossing close to the source edge, e.g. in a non-transiting event, the magnification peak maximizes in the $I$-band.

In addition to the magnification peak, the width of the lightcurve depends on the passband filter too. The Full Width at Half Maximum of a microlensing lightcurve ($\Delta_{\rm t}$) is given by
\begin{eqnarray}
\Delta_{\rm t}= 2 t_{\rm E} \sqrt{D-u_{0}^{2}}, 
\end{eqnarray}
\noindent where $$D=\frac{4-A_{\rm{m}}^{2} + A_{\rm{m}}\sqrt{A_{\rm{m}}^{2}-4}}{A_{\rm{m}}^{2}/2-2}.$$ $D$ decreases with increasing value of the magnification peak. Therefore, in transit microlensing events, the magnification peak decreases and its width increases (i.e. the lightcurve is more flattened) in longer wavelengths. These two factors make the chromatic perturbations of LD effects on microlensing lightcurves be symmetric with respect to the time of closest approach, $t_{0}$, and also maximize the perturbation at the time of closest approach. 

The LD-induced chromatic perturbations are highlighted during CC features as well. We note that during caustic crossings these chromatic perturbations have the same tendency as during the high magnification peak (see the residual part of \ref{fig2d}), but during caustic crossing the perturbations are not symmetric (with respect to the time the source center passes the caustic curve), unless the source trajectory passes from the binary axis normally. The maximum perturbation happens when the source edge is on the caustic curve and the source center is out of it (i.e. while the magnification factor is sharply increasing). At that time, the magnification factor at longer wavelengths is higher than at shorter wavelength.

The time scale of the LD-induced chromatic perturbations is of order $t_{\ast}=t_{\rm E}~\rho_{\ast}$, i.e., the source radius crossing time. From simulations, we note that the events with the shortest $t_{\ast}$ are least suitable for measurements of their chromatic perturbations.
	
\noindent The time scale of BL-induced chromatic deviations is of order $t_{\rm E}$ which is longer than $t_{\ast}$. Additionally, these BL-induced deviations remain approximately constant while the lens is crossing the source disk. As a result, in the residual parts of Figures \ref{three}, the difference between solid and dashed curves are highlighted only when either the lens is crossing the source disk or the source disk is passing the caustic lines. So, the BL-induced chromatic deviations can be subtracted from the chromatic curves. That helps us to study the chromatic deviations due to any possible perturbations in the stellar brightness profiles. In this sense, the BL effect will not make degenerate chromatic perturbations with the LD effect.

For transit microlensing events, the chromatic perturbations due to only the LD effect can reach to $0.025$-$0.04$ mag close to the peak of the light curve, which is detectable with LI camera at the Danish $1.54$ m telescope. For non-transit HM (single) microlensing events in which the lens passes very close to the source edge $u_{0}\simeq \rho_{\ast}$, the LD-induced chromatic perturbation at the time of the closest approach can reach $\sim 0.01$ mag, again detectable with the LI camera (see Figure \ref{fig2b}). 

\noindent During CC features the LD-induced chromatic perturbations can reach to $0.05$-$0.1$ mag (see Fig. \ref{fig2c}), but the duration of such high signal is generally too short (it happens when the source edge is entering the caustic and its center is out of it). If the source star is moving parallel with the caustic curve or when the source size is comparable with the caustic size, the time scale of the chromatic perturbations increases and their chromatic perturbations can be measured. Another way to detect such short-duration perturbations is improving the observing cadence. From our simulation of HM and CC microlensing events, we note that the LD-induced perturbations in $B$-$R$ are always larger than those in $V$-$I$.   

We perform a Monte-Carlo simulation from HM and CC microlensing events from limb-darkened and blended source stars toward the Baade's window. We find that for the HM events, the probabilities of detecting the chromatic perturbations with low and high sensitivities are $62\%$ and $50\%$, respectively. For CC binary microlensing events, the probabilities of detecting the chromatic perturbations with the LI camera are $31\%$ (low sensitivity) and $22\%$ (high sensitivity) in each CC feature (as reported in the rows of Table \ref{tab_result} which are labeled by BL \& LD). Generally, during CC, magnification factors are not as high as for HM microlensing events. That causes larger error bars for data taken during CC than those taken in HM events. The dependence of error bars to the stellar apparent magnitudes is shown in Figure \ref{errordd}. 
\begin{figure*}
\centering
\subfigure[]{\includegraphics[angle=0,width=0.49\textwidth,clip=0]{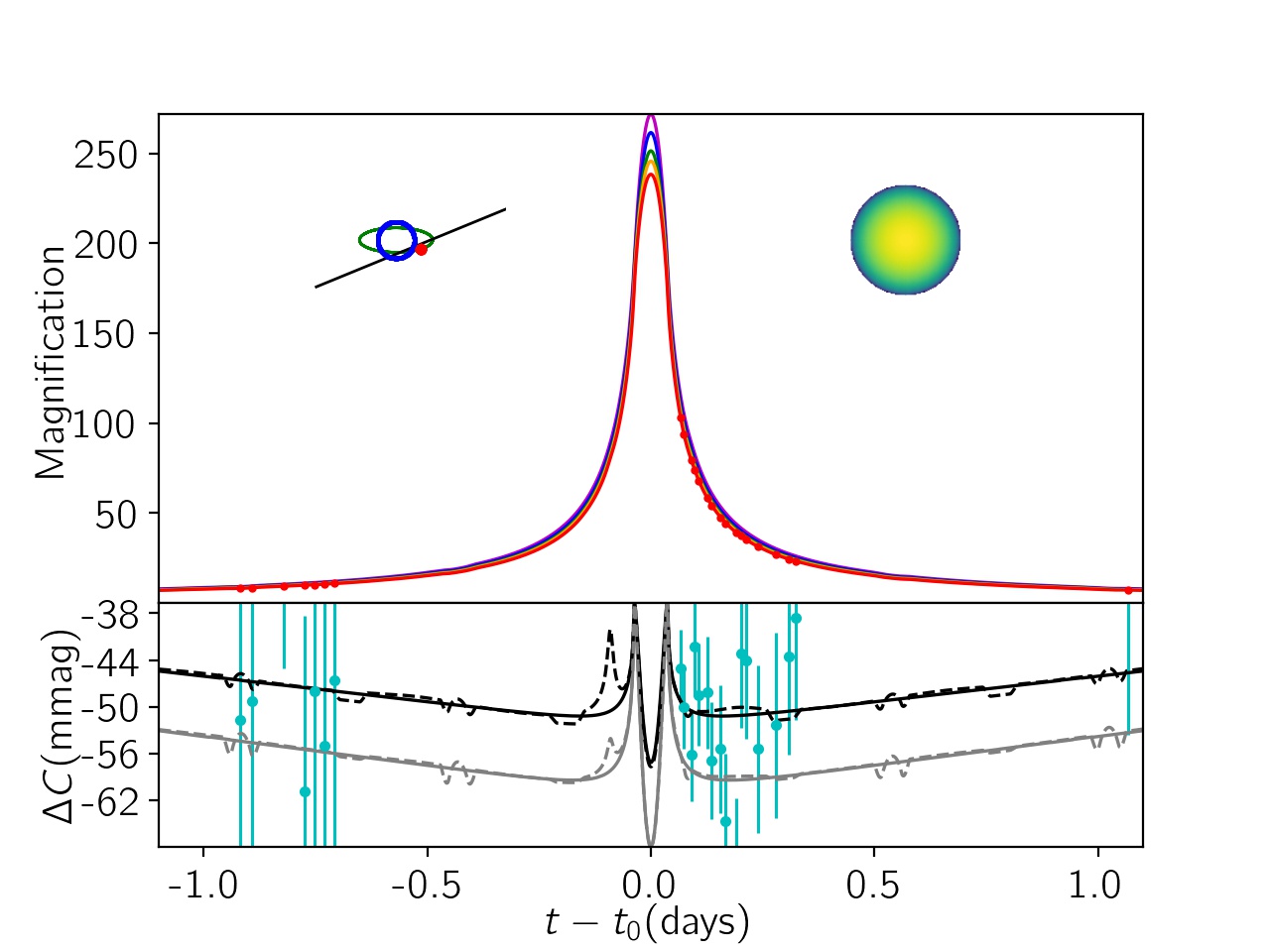}\label{fig5a}}
\subfigure[]{\includegraphics[angle=0,width=0.49\textwidth,clip=0]{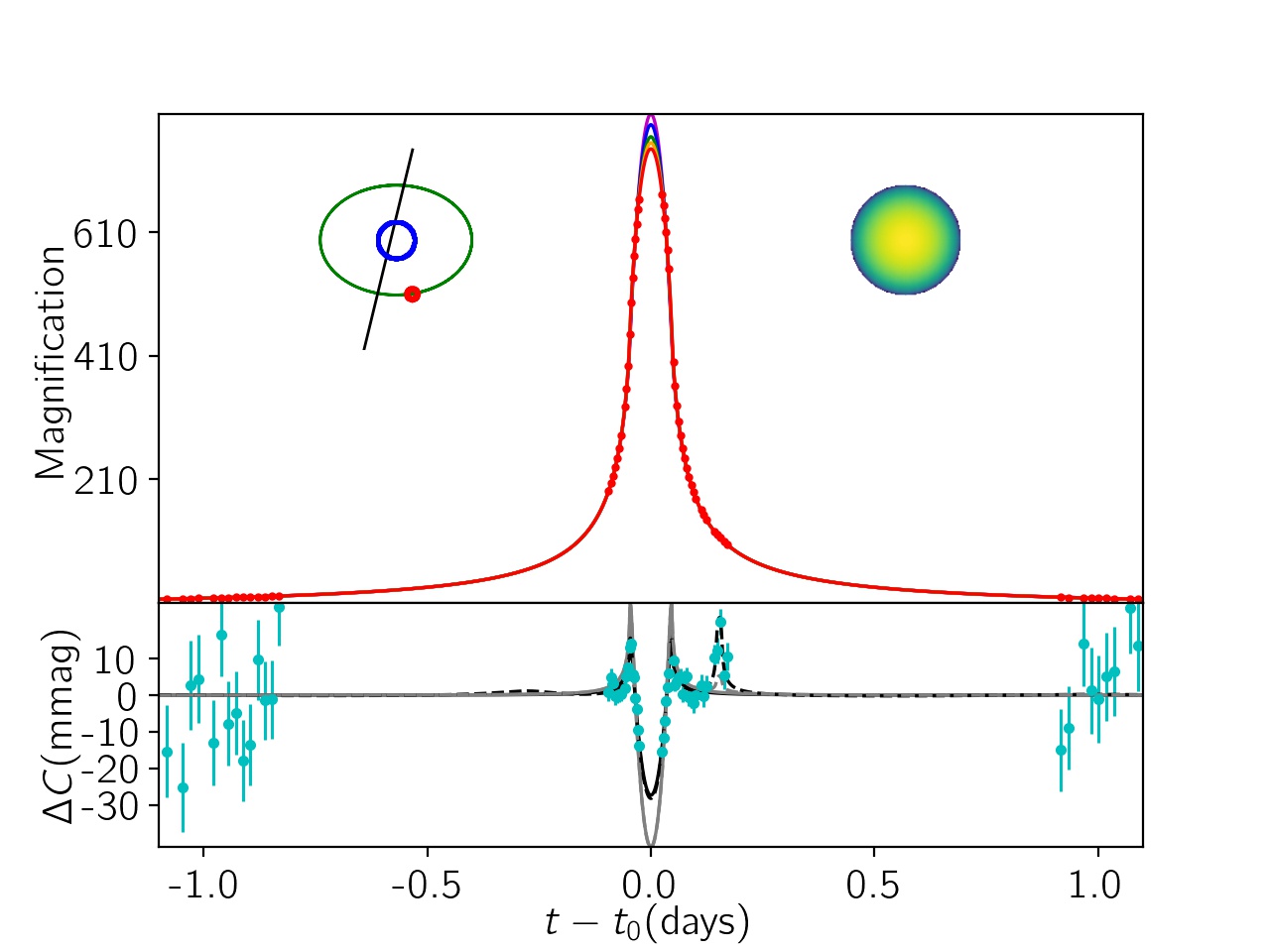}\label{fig5b}}
\subfigure[]{\includegraphics[angle=0,width=0.49\textwidth,clip=0]{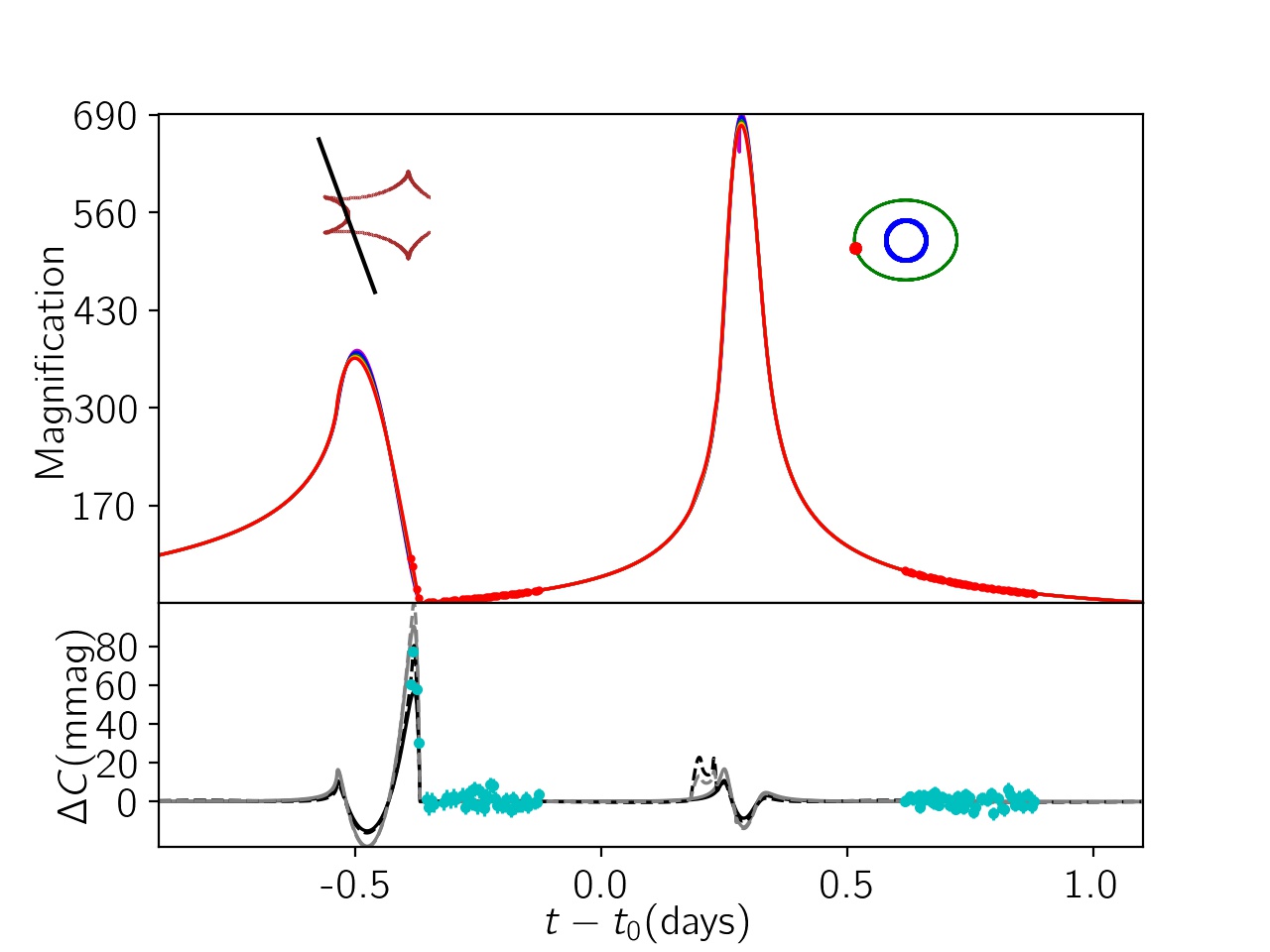}\label{fig5c}}
\subfigure[]{\includegraphics[angle=0,width=0.49\textwidth,clip=0]{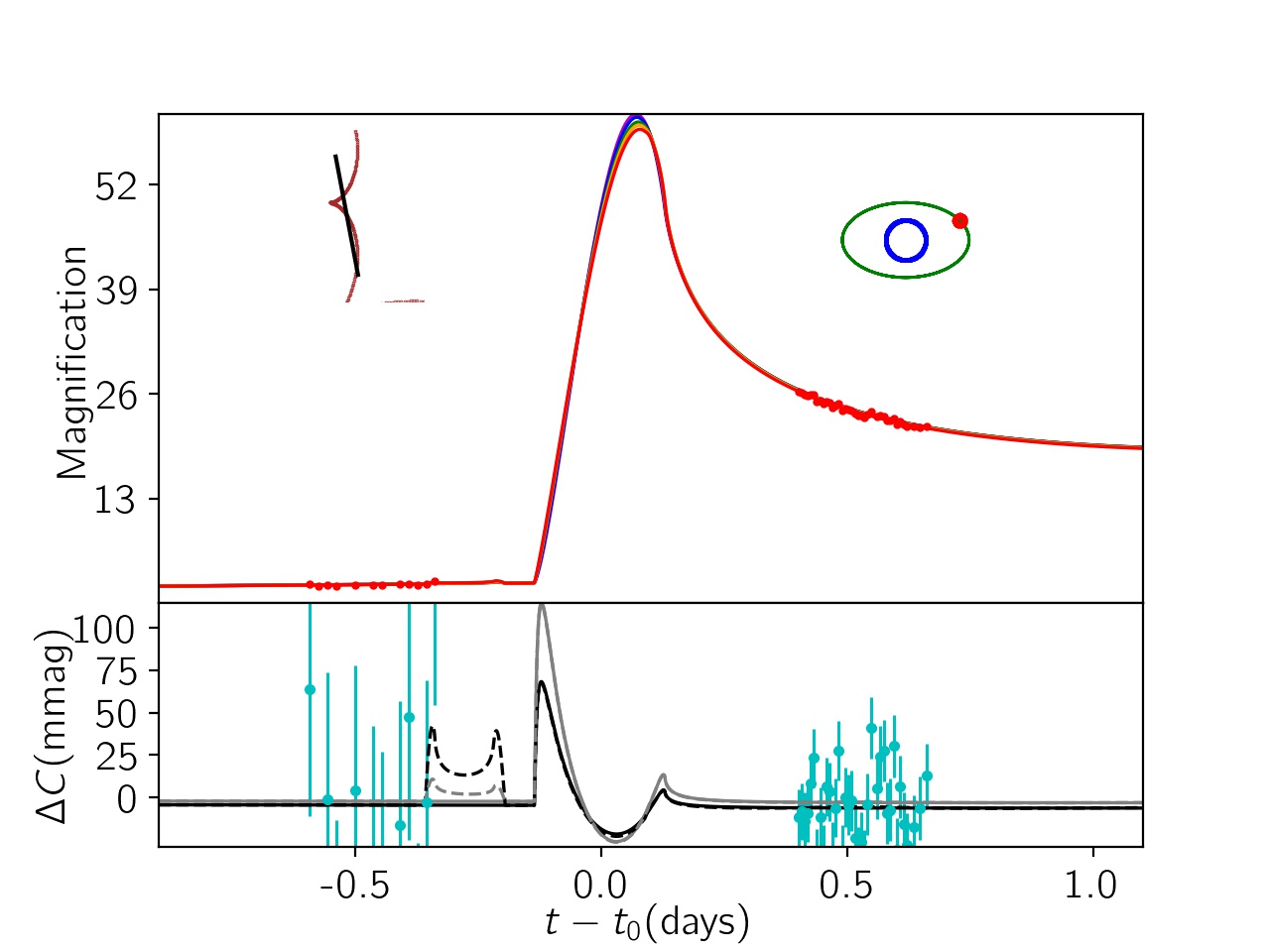}\label{fig5d}}
\caption{Similar to Figure \ref{three}, but we additionally consider a close-in giant planet around each (blended and limb-darkened) source star. In each panel, the source star (blue circle), its planet (red circle), the planet orbit projected on the sky plane (green curve) and the lens trajectory (black solid line) are shown as insets in the figures. The chromatic perturbations in the $V$-$I$ and $B$-$R$ stellar colors by considering only the LD and BL effects (solid) and LD, BL plus a source star planet (dashed) are shown in the residual parts with black and gray curves. Their parameters are given in Table \ref{tab2}.}\label{last}
\end{figure*}
\begin{table*}
\caption{The parameters of the lightcurves shown in Figure \ref{last}.}             
\label{tab2}      
\centering          
\begin{tabular}{cccccccccccccc}    
\toprule[1.5pt]
$~$&$\log_{10}[u_{0}]$&$t_{\rm{E}}$&$\log_{10}[\rho_{\ast}]$&$m_{\rm{base},~I}$&$T_{\rm{eff}}$ 	&$a$&$T_{\rm{p}}$&$i$& $b_{I}$&$q$&$d$&$\xi$&$\Delta \chi^{2}$\\
&&$\rm{(days)}$&&$\rm{(mag)}$&$\rm{(K)}$&$\rm{(au)}$&$\rm{(K)}$&$\rm{(deg)}$&&&&$\rm{(deg)}$&\\ 
\toprule[1.5pt]
\ref{fig5a}& -2.35 & 8.7 &  -2.20 &  18.7 & 5846 & 0.01 & 2918 & 70& 0.86 & - & - & 202 & 747 \\
\ref{fig5b}& -3.10 &17.7 &  -2.55 & 18.1 & 5527 & 0.02 & 2175 & 43 & 1.0&  - & - & 76 &  876\\
\ref{fig5c}&-2.02 & 82.4 &  -3.27 & 17.3 & 7480 & 0.02& 3736 & 38 &1.0& 0.01 & 1.09 & 110  & 1875 \\
\ref{fig5d}&-0.63 & 31.9 &  -2.80 & 19.3 & 6055 & 0.01 & 2549 & 53 & 0.99 & 0.44 & 1.21 & 100 &  6 \\
\hline
\end{tabular}
\tablefoot{Here, $a$ is the radius of the planet orbit, $T_{\rm p}$ is the planet temperature, $i$ is the inclination angle of the planetary orbit with respect to the sky plane.}
\end{table*}

Accordingly, if the Danish $1.54$ m telescope in each observing season covers the magnification peaks of $\sim 10$ HM microlensing events (with $u_{0}<0.01$) and $\sim 10$ CC features in binary events during two-day time intervals around their peaks \footnote{For instance, the Danish telescope has densely followed $68$ anomalous and highly-magnified microlensing events alarmed by the ARTEMiS system in 2008 with high photometric precisions \citep{Dominik2010}.}, we can therefore expect to be able to detect the chromatic perturbations in $\sim 7$ events per observing season even within our standard priority of events.

Generally, the brighter source stars in longer microlensing events are the most suitable candidates for measuring their chromatic perturbations (see, Fig. \ref{fig2a}), e.g., the events of red giant source stars with large $\rho_{\ast}$ values. These stars are intrinsically bright and because of their large radii, their $t_{\ast}$ values increase. In out simulations, we in fact used the emergent radiations for cool stars with $\log[g]\geq 3.0$ as developed by P.~Harrington, corresponding to main-sequence stars. For red giant stars, the LD effect is stronger \citep[see, e.g., ][]{2011Zub}, which makes them the most suitable candidates for chromatic measurements.

An important conclusion from our simulations is that if there are several sets of data points for a given microlensing event taken by different telescopes with different filters, then for these sets of data the resulting lightcurves will be slightly different because the related LD coefficients and blending parameters are different. Considering the same LD coefficients for all data sets generates some extra noises in the data (around the peaks) with respect to the best-fitted model. We note that the conversion of the magnitudes of the different telescopes and filters to a common filter system (before finding the best-fitted model) does not solve this problem, because the discrepancy is time-dependent between the magnified magnitudes in different filters, whereas the conversion relations for the magnitudes in different filters are applicable to the unperturbed baseline data.  

In the next section, we study another source for chromatic perturbations which is Close-in Giant Planet Companions around the microlensing source stars.  

%%%%%%%%%%%%%%%%%%%%%%%%%%%%%%%%%%%%%%%%%%%%%%%%%%%%%%%%%%%%%%%%%%%%
\section{Close-in Planetary Systems}\label{CGPs}
Close-in Giant Planet Companions (CGPs), or the so-called hot Jupiters, have been discovered abundantly during the Kepler mission and through transit or radial velocity measurements from the ground \citep{2010Kepler}. Hot Jupiters are Jupiter-like planets located at the distance range $[0.01,~0.1]$ au from their parent stars. Their orbits are most often tidally locked and their orbital periods are $\sim3$-$10$ days. These objects perturb their host star brightness but also their color. These faint and small color distortions are best observable at long wavelengths, e.g. infrared, in HM or CC microlensing events through photometric or polarimetric observations \citep{Graff2000, Sajadian2010Jupi, sajadian2017pol}. This channel of exoplanet detection is parallel with the main channel of planet detection in microlensing events where the gravitational effect of a planet in the lensing system affects the light path of an image of the source star \citep[see, e.g., ][]{MaoP1991, 2006Natur5.5, gaudi2012rev}. In this section, we study how CGPs orbiting the source star can make detectable chromatic perturbations in HM and CC microlensing events. 

For simulating close-in planetary systems as microlensing sources, the following assumptions are applied. (i) The planet orbit is circular (its eccentricity is zero).  (ii) The planet temperature is uniform over its surface. (iii) Each source star is orbited by a CGP (such that the statistics becomes per CGP). (iv) The source star acts as a black body and its radiation is thermal and specified according to its effective temperature.

\noindent For projection of the planetary orbit onto the plane of the sky, we use an inclination angle $i$ which is the angle between the orbital plane and the plane of the sky. If $i=0$ the planetary orbit will be circular as seen by the observer. The phase angle, $\beta$, i.e., the angle between line of sight to the observer and the host source star as seen by the planet, indicates the position of the planet with respect to the source center. If $90^{\circ}<\beta<180^{\circ}$, the planet is transiting in front of the source surface and can block the source brightness and for $0<\beta<90^{\circ}$ the planet is passing behind its host star and its light can be blocked by host star. To estimate the planet radiation, we consider two components, thermal and reflected ones. All details about calculating these radiations can be found in \citet{sajadian2017pol}.

In Figure \ref{last}, four examples of HM and CC microlensing events of close-in planetary systems are shown. The insets represent the planetary orbits projected onto the sky plane (green curves), the host stars and planets (blue and red circles) and the lens trajectory (black solid lines). For the host stars we consider the LD and BL effects as well. The chromatic perturbation in the source colors by considering only LD and BL effects are shown by solid (black and gray) curves in the residual parts. The total chromatic perturbations due to the source stars and their CGPs are shown by dashed (black and gray) curves. The parameters of these lightcurves are reported in Table \ref{tab2}. Some key points from these lightcurves are mentioned in the following.\\

\begin{itemize}
\item The CGPs make small extra perturbations in the stellar color curves which mostly happens when the lens is passing very close to planets, e.g., Figures \ref{fig5a} and \ref{fig5b}. 
	
\item The planet-induced perturbations are strongest in the $I$-band. Hence, the planet-induced perturbations are larger in the stellar color $V$-$I$ than $B$-$R$, e.g., see Figure \ref{fig5b}. 
	
\item The edge-on planetary orbits, with the inclination angle $i\sim 90^{\circ}$, make larger chromatic perturbations, i.e., the main source for chromatic perturbation for the source star in these systems is the occultation of the stellar brightness by the planet transit.  	
\end{itemize}

\noindent By assuming $\Delta \chi^{2}>100,~200$ as detectability criteria with low and high sensitivities, we conclude that the detection probabilities for chromatic perturbations in HM microlensing of close-in planetary systems is $64\%$ and $53\%$, respectively. We note that by considering only LD and BL effects for the source star this probability is $50$-$62\%$, i.e., CGPs improve the detection probability of chromatic perturbations by $\sim 2\%$. In binary CC microlensing events, the probability of detecting CGPs is $23$-$32\%$ and reported in Table \ref{tab_result}.

\noindent In the next section, we consider one stellar spot for each source star which are highly magnified and study the characteristic and statistic of the resulting chromatic perturbations.

\begin{figure*}
\centering
\subfigure[]{\includegraphics[angle=0,width=0.49\textwidth,clip=]{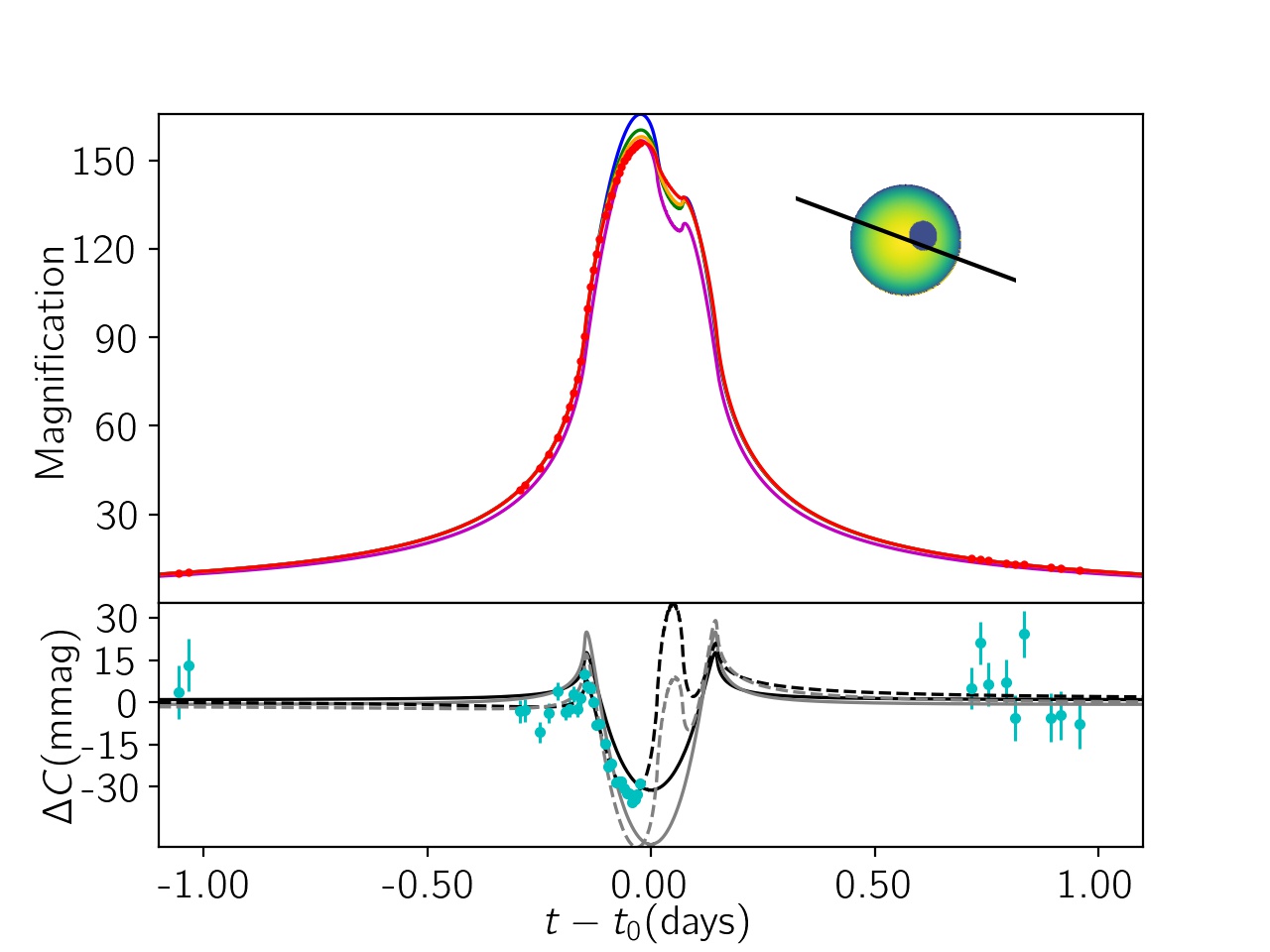}\label{fig3a}}
\subfigure[]{\includegraphics[angle=0,width=0.49\textwidth,clip=]{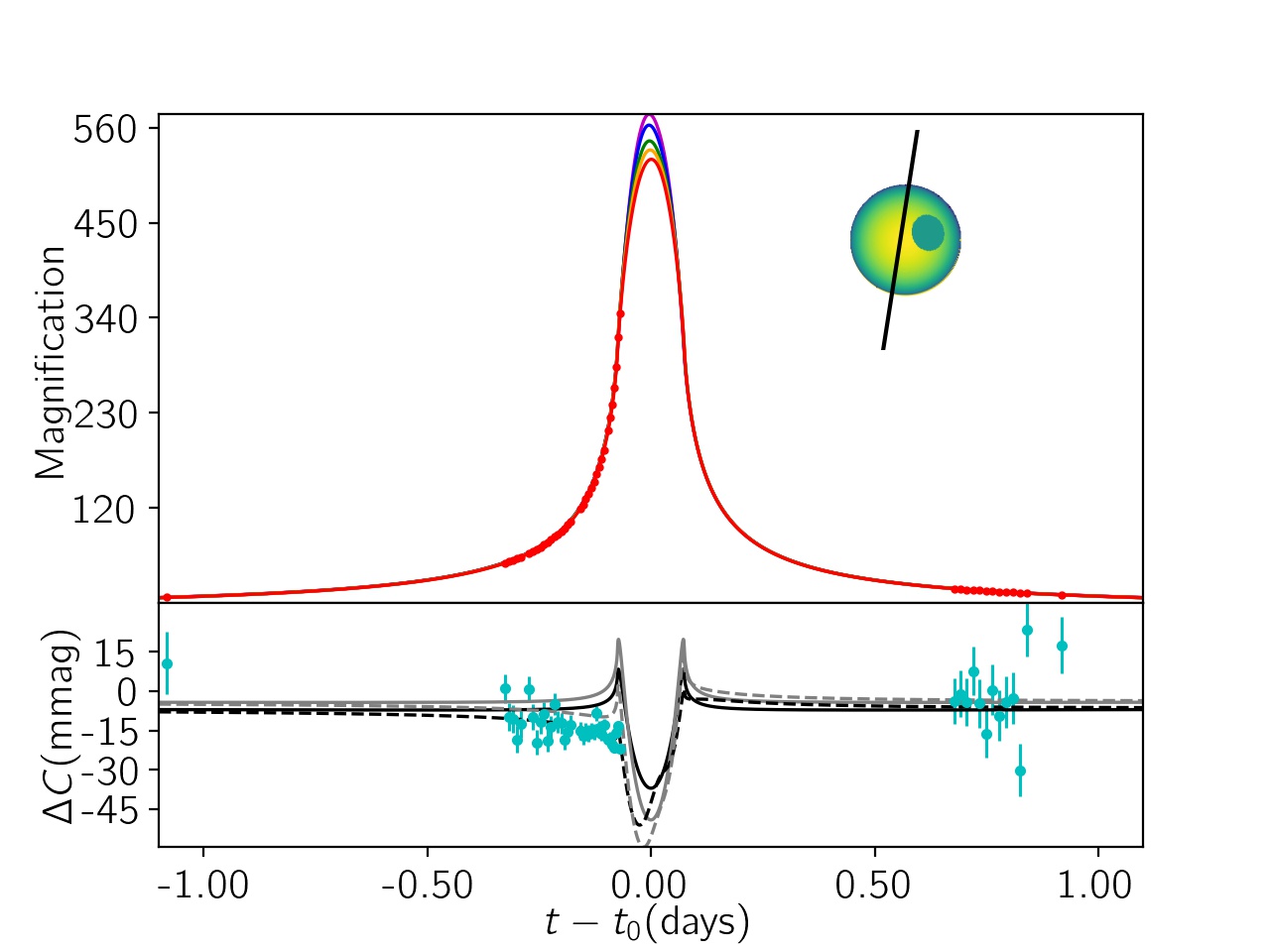}\label{fig3b}}
\subfigure[]{\includegraphics[angle=0,width=0.49\textwidth,clip=]{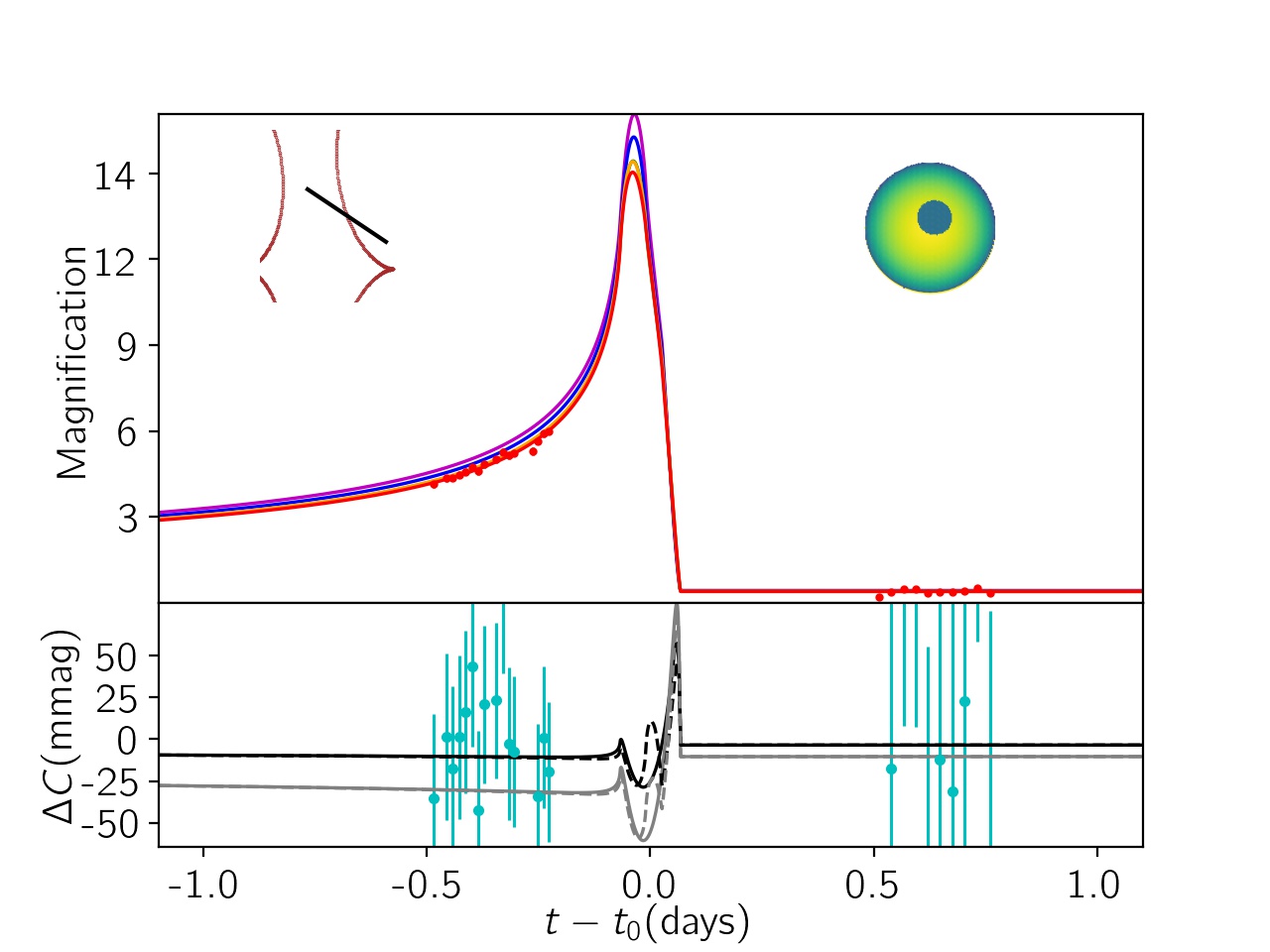}\label{fig3c}}
\subfigure[]{\includegraphics[angle=0,width=0.49\textwidth,clip=]{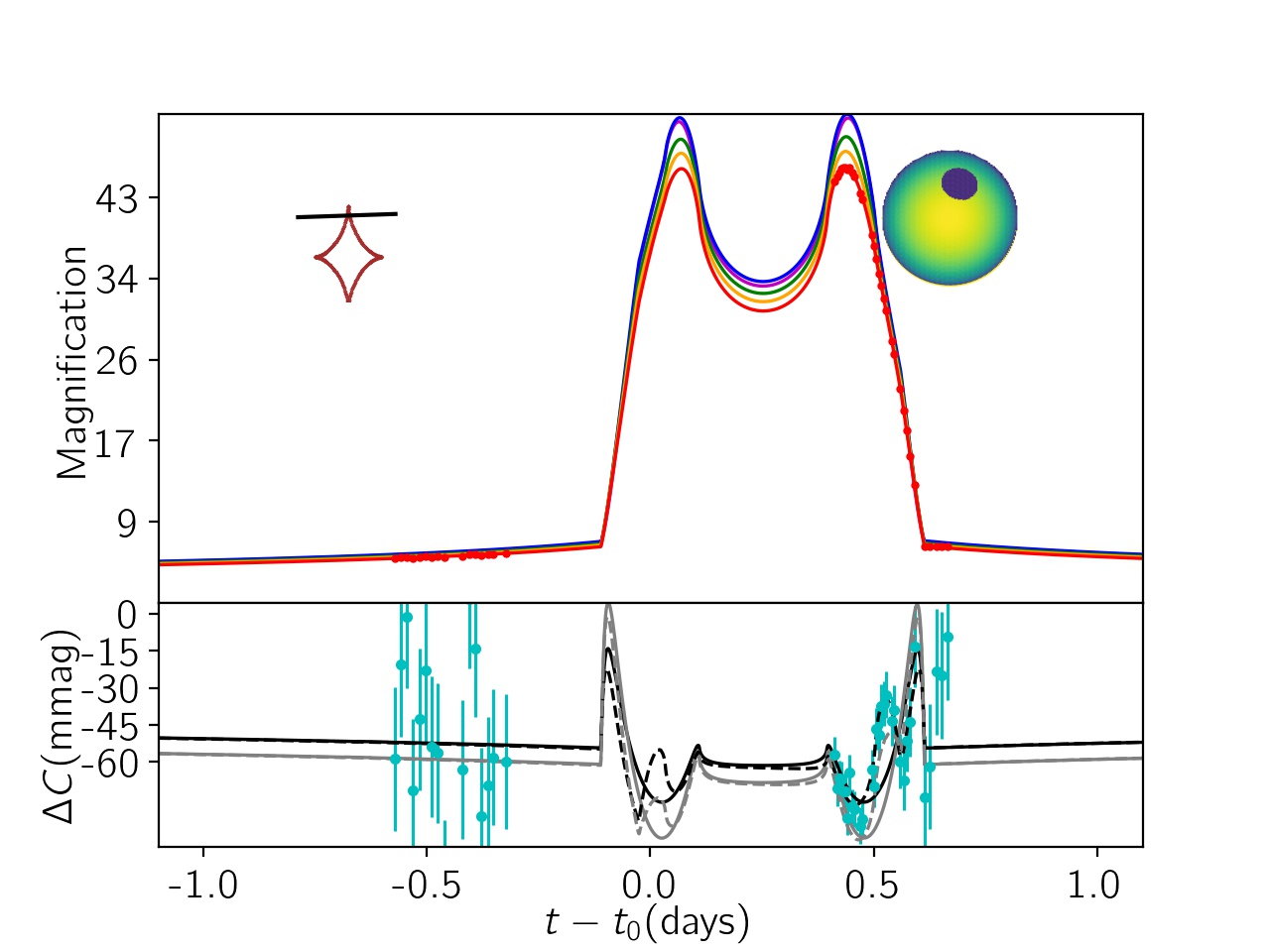}\label{fig3d}}
\caption{Similar to Figures \ref{three} and \ref{last}, but for the spotted source stars. In these plots, the brightness profiles over the source disks (by considering LD and stellar spots) and the lens trajectory (solid black lines) are shown in the right-hand inside parts. The chromatic perturbations in the $V$-$I$ and $B$-$R$ stellar colors by considering only LD, BL effects (solid) and LD, BL as well as stellar spots (dashed) are shown in the residual parts with black and gray curves. The parameters of the lightcurves are reported in Table \ref{tab3}.}\label{four}
\end{figure*}
\begin{table*}
	\caption{The parameters of the lightcurves shown in Figure \ref{four}.}             
	\label{tab3}      
	\centering          
	\begin{tabular}{ccccccccccccc}    
		\toprule[1.5pt]
		$~$&$\log_{10}[u_{0}]$&$t_{\rm{E}}$&$\log_{10}[\rho_{\ast}]$&$m_{\rm{base},~I}$&$T_{\rm{eff}}$	&$\theta_{0}$&$f_{V}$& $b_{I}$&$q$&$d$&$\xi$&$\Delta \chi^{2}$\\ &&$\rm{(days)}$&&$\rm{(mag)}$&$\rm{(K)}$&$\rm{(deg)}$&&&&&$\rm{(deg)}$&\\ 
		\toprule[1.5pt]
		\ref{fig3a}& -3.78 & 11.6 &  -1.89 &  17.3 & 5726 & 15.76 & 0.35 & 0.94 & - & - & 340 & 3774\\
		\ref{fig3b}& -3.41 & 18.7 &  -2.40 &  18.3 & 5860 & 19.41 & 0.61 & 0.99 & - & - & 81 & 956\\% 26th Aug
		\ref{fig3c}& -0.57 & 49.4 &  -3.09 &  19.3 & 5394 & 15.54 & 0.44 & 0.38 & 0.78 & 0.95 & -34 & 0\\% 26th Aug
		\ref{fig3d}& -0.77 & 48.9 &  -2.66 &  18.4 & 5889 & 15.84 & 0.37 & 0.83 & 0.98 & 0.56 & 2 & 1672\\%%26th Aug
		\hline
	\end{tabular}
\tablefoot{Here, $f_{V}$ is the ratio of the stellar brightness over the spot to its amount at the source center in $V$-band without spot, and $\theta_{0}$ is the enclosing angle to the stellar spot as measured from the source center.}
\end{table*}
\section{Stellar spots}\label{spots}
The effects of stellar spots (SS) in microlensing lightcurves have been studied extensively in several references. Here, a brief review of them is offered. Detecting stellar spots in single and binary microlensing events was studied by \citet{Heyrovski2000, Han2000spot} and they reported the spot-induced deviations for spots with radii smaller than $0.2 R_{\ast}$ can reach up to $2\%$. Detecting stellar spots in different wavebands as a function of the spot temperature was also investigated in \citet{Hendry2002}. Additionally, \citet{2010Hwang} has also studied the detection and characterization of stellar spots in the next generation of microlensing observations with short cadence and concluded the size and the location of the stellar spots are measurable through such observations. Polarimetry follow-up observation from HM microlensing events of red giant source stars is an alternative approach for illuminating stellar spots and even measuring the stellar magnetic field. In that case, stellar spots perturb the polarimetry curves \citep{2015sajadian}. Intrinsic rotation of the source star around its axis causes the location of its stellar spot to vary while the lens is transiting the source surface in HM microlensing events. Accordingly, stellar rotation alters the spot-induced perturbations on microlensing lightcurves \citep{2015Giordano}. 

Here, we study the chromatic perturbations due to stellar spots in HM and CC microlensing events. For simulating stellar spots we follow the approach introduced in \citet{2015sajadian} and consider four parameters. $f_{V}$ is the ratio of the stellar brightness over the spot to its amount at the source center in $V$-band without spot, $\theta_{0}$ is the enclosing angle to the stellar spot as measured from the source center and with respect to the axis passing from the spot center. It determines the spot area. We also need two angles, $\xi$ and $\phi$ to project the stellar spot on the sky plane. In the projection process, there are two coordinate systems: the stellar coordinate system $(x_{\ast},~y_{\ast},~z_{\ast})$ and the observer one $(x_{\rm o},~y_{\rm o},~z_{\rm o})$. We set the spot on the stellar pole (in the stellar coordinate system) with $\theta_{\ast}<\theta_{0}$. One can convert the first coordinate system to the second one by two consecutive rotations, around $y_{\ast}$-axis by the angle $\xi$ and then around $z_{\ast}$-axis by the angle $\phi$, as: 
\begin{eqnarray}
x_{\rm o}&=& \cos \phi \cos \xi~x_{\ast} - \sin \phi~y_{\ast} + \cos \phi \sin \xi~z_{\ast}\nonumber\\
y_{\rm o}&=& \sin \phi \cos \xi~x_{\ast} + \cos \phi~y_{\ast} + \sin \phi \sin \xi~z_{\ast}\nonumber\\
z_{\rm o}&=& -\sin \xi~x_{\ast} +\cos \xi~z_{\ast}
\end{eqnarray}

\noindent The points with $z_{\rm o}>0$ are detectable by the observer and make the projected stellar disk on the sky plane. We consider only one spot for each source star. For each spot the enclosing angle is chosen uniformly from the range $[5,~20]$ deg. The fraction $f_{V}$ is also taken randomly as $f_{V}\in [0.1,~0.7]$. Two projection angles are selected randomly from the range $[5,~85]$ deg. We consider the LD and BL effects for each source star and add one spot in a random place over the source surface.

\noindent The overall magnification factor of the spotted source star in the adopted filter, $A_{\mathcal{S},~F}$, can be given by:  
\begin{eqnarray}
A_{\mathcal{S},~F}(t)=\frac{\int_{\ast}dx_{o} dy_{o}~A(t)~I_{\ast,~F} +\int_{s} dx_{o} dy_{o}~A(t)~(I_{\rm s,~F}-I_{\ast,~F})}{\int_{\ast}dx_{o} dy_{o}~I_{\ast,~F} +\int_{\rm s} dx_{o} dy_{o}~(I_{\rm s,~F}-I_{\ast,~F})},
\end{eqnarray}

\noindent  where the indexes $s$ and $\ast$ refer to the disks of the spot and the source star (both projected on the sky plane), respectively.  $I_{\ast,~F}$ and $I_{\rm s,~F}$ are the intensities of the source star considering the LD effect and its spot in the fiter $F$, respectively.  $A(t)$ is the magnification factor of each element with the coordinate $(x_{o},~y_{o})$. If we assume constant amounts for the intensities, $I_{\ast,~F}$ and $I_{\rm s,~F}$, and ignore the LD effect, the overall magnification factor can be simplified as following:  

\begin{eqnarray}\label{atot}
A_{\mathcal{S},~F}(t)= \frac{A_{\ast}(t)-A_{\rm s}(t)~\alpha~(1-f_{F})}{1- \alpha~(1-f_{F})},
\end{eqnarray}

\noindent where,  $A_{\ast}(t)$ and $A_{\rm s}(t)$ are the magnification factors of the whole source star and its spot, respectively, $\alpha$ is the ratio of the spot area to the source area (both projected on the sky plane).  We note that although $f_{F} \in [0,~1]$, $\alpha$ has very small values, which makes the factor $\alpha (1- f_{F})$ be very small. $f_{F}$ depends on the waveband and therefore generates SS-induced chromatic perturbations in HM and CC microlensing events, even if we ignore the LD effect of the source star. By adding the BL effect, the magnification factor is:  
\begin{eqnarray}
A_{b,~\mathcal{S},~F}= b_{F}~A_{\mathcal{S},~F}~+ ~1- b_{F}.
\end{eqnarray} 
 
Four examples of the microlensing lightcurves from spotted source stars are shown in Figure \ref{four}. The parameters of these lightcurves are given in Table \ref{tab3}. For the source stars the LD and BL effects are considered as well. In the residual parts, we plot the perturbations in the source color $V$-$I$ (black) and $B$-$R$ (grey) due to the LD and BL effects (solid curves) and LD, BL plus stellar spots (dashed curves). The spot size depends on both projection angles and the enclosing angle. Right-hand insets represent the brightness profiles of the spotted source star and the lens trajectories projected on the source plane (solid black lines). We note the following key points from the figures:

\begin{itemize}
\item Spots make asymmetric chromatic perturbations with respect to time of the closest approach, unless the spot is exactly on the source center (which is rare) or the lens-spot distance minimizes at $t_{0}$. We note that the LD and BL effects itself causes symmetric perturbations.  
	
\item If the lens is crossing the spot disk, then generally the whole magnification decreases, but at the same time the spot acts as an extra, faint and cooler source star, which affects the $I$-band, so that the spot significantly perturbs the lightcurve in the $I$-band filter. In that case, a significant perturbation is made in the $V$-$I$ color, whereas the LD effect of the source star is mostly higher in $B$-$R$ band, see Fig. \ref{fig3a}. In these cases (i.e. when the lens transits the spot surface), the chromatic perturbations in the $V$-$I$ stellar color can reach $0.07$ mag. 
	
\item If the lens passes close to the stellar spot but not over its surface, the spots make perturbations in the chromatic curves due to limb-darkened and blended source stars. These perturbations for the spots located at the source edges are too short to be observed.  
	
\item Generally, the spot makes the chromatic perturbations in $V$-$I$ and $B$-$R$ have different forms and their symmetry with respect to the time of the closest approach breaks.  
	
\item As seen from Figure \ref{fig3a}, the spot-induced perturbations in microlensing lightcurves can be similar to planetary ones. In that case, one possible test to resolve these degenerate solutions is measuring the stellar color. If it changes with time, the perturbation source should be a stellar spot, otherwise it is likely to be a planet around the lens object.    
	
\item If the position of the stellar spot is projected on the center of the source disk, it can make similar and degenerate magnification curves and chromatic perturbations to that made by a transiting giant planet. Because, they make similar perturbations for the stellar brightness profile. 

\item The time and duration of the perturbation depends on the spot location. If the spot is on the stellar edge, its size after projection on the source disk will usually be very small which will result in a very short-duration perturbation. We note that, if  the observing cadence is improved during caustic crossing, the probability for detecting such perturbations increases significantly. 
\end{itemize}

In order to study the detectability of the chromatic perturbations in HM microlensing events of the limb-darkened, blended, and spotted source stars, we have simulated an ensemble of these events and assume that when $\Delta \chi^{2}>100,~200$ the chromatic perturbations are detectable with low and high sensitivities. We conclude that these perturbations are measurable in $\sim 62.7,~51.0\%$ of the events when observed by the LI camera near their peaks. One example is shown in Figure \ref{fig3a}.  In binary CC events, this probabilities are  $36,~27\%$. 

\noindent According to the results of the simulation in the previous section, the spot-induced perturbations improve the detectability by $\sim 0.5,~5.0\%$ in HM and CC events, respectively. The stellar spots most often pass from the caustic curves during CC features (see, bottom panels of Fig. \ref{four}), whereas the probability of crossing the spot disk by the lens in HM events is low.

In the next section, we assume the microlensing source stars are hot and early-type ones with gravity-darkening effects and study their chromatic perturbations.

%%%%%%%%%%%%%%%%%%%%%%%%%%%%%%%%%%%%%%%%%%%%%%%%%%%%%%%%%%%%%%%%%%%%%%%%%%%
\begin{figure*}
\centering
\subfigure[]{\includegraphics[angle=0,width=0.49\textwidth,clip=0]{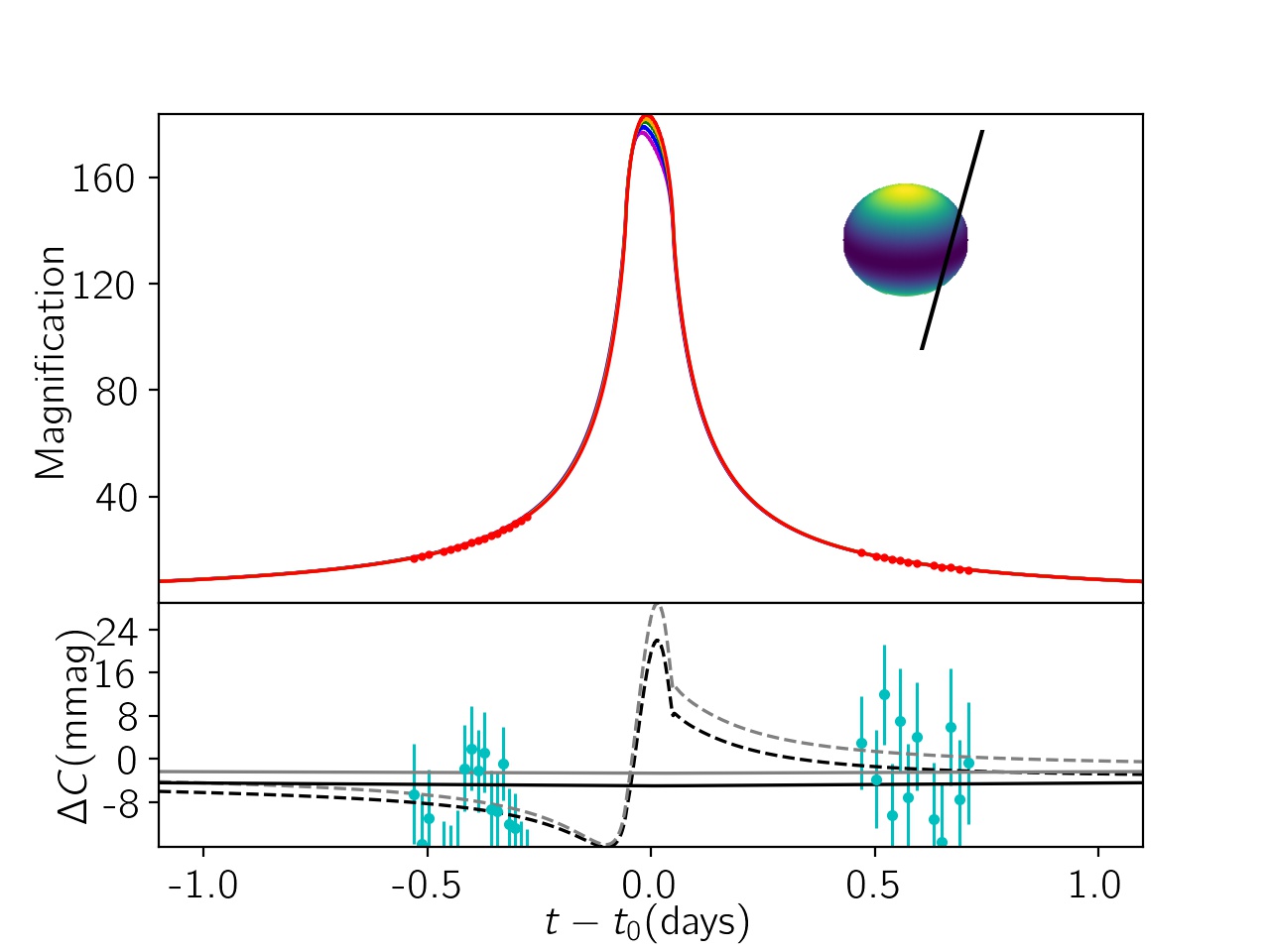}\label{fig4a}}
\subfigure[]{\includegraphics[angle=0,width=0.49\textwidth,clip=0]{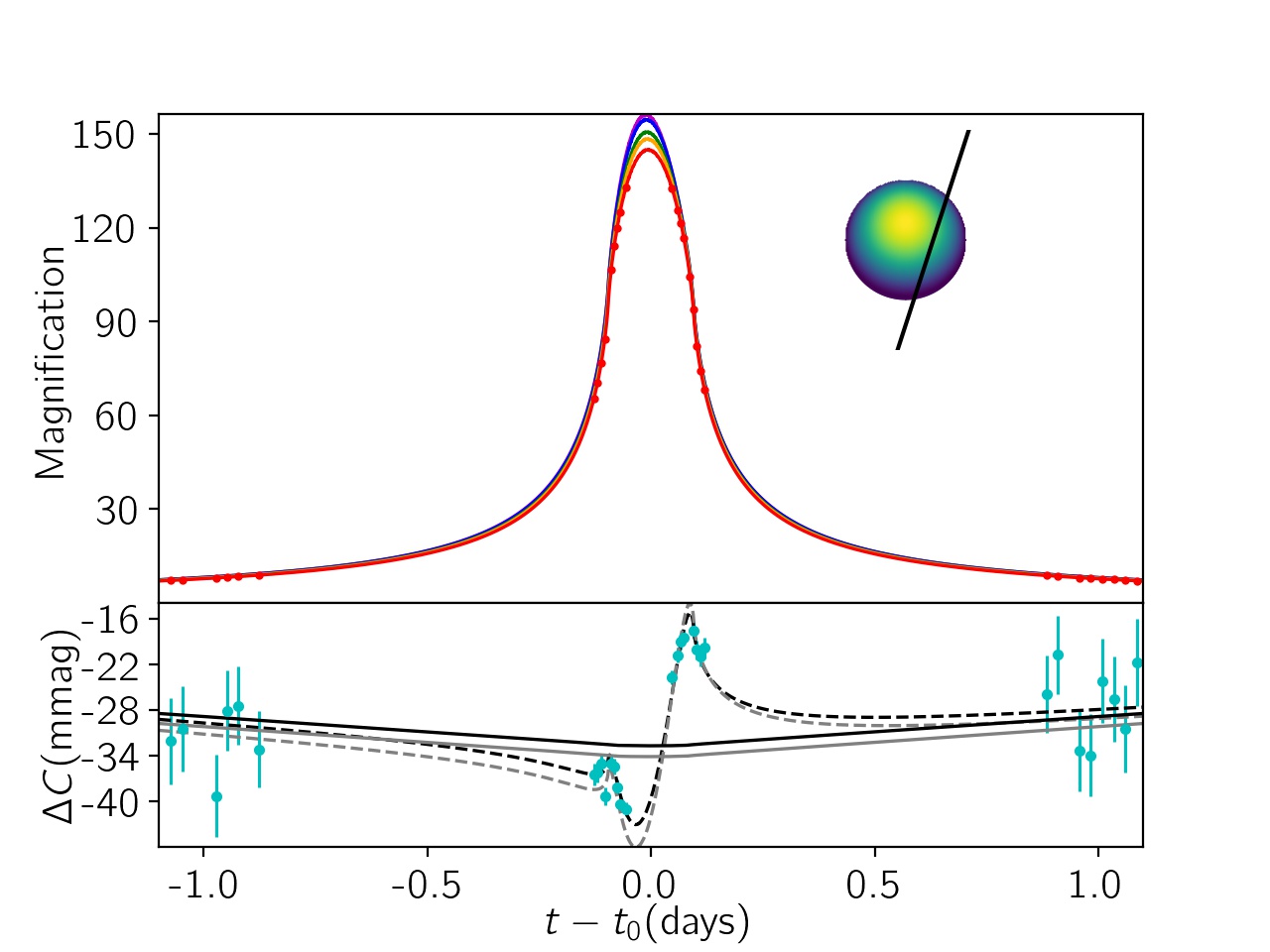}\label{fig4b}}
\subfigure[]{\includegraphics[angle=0,width=0.49\textwidth,clip=0]{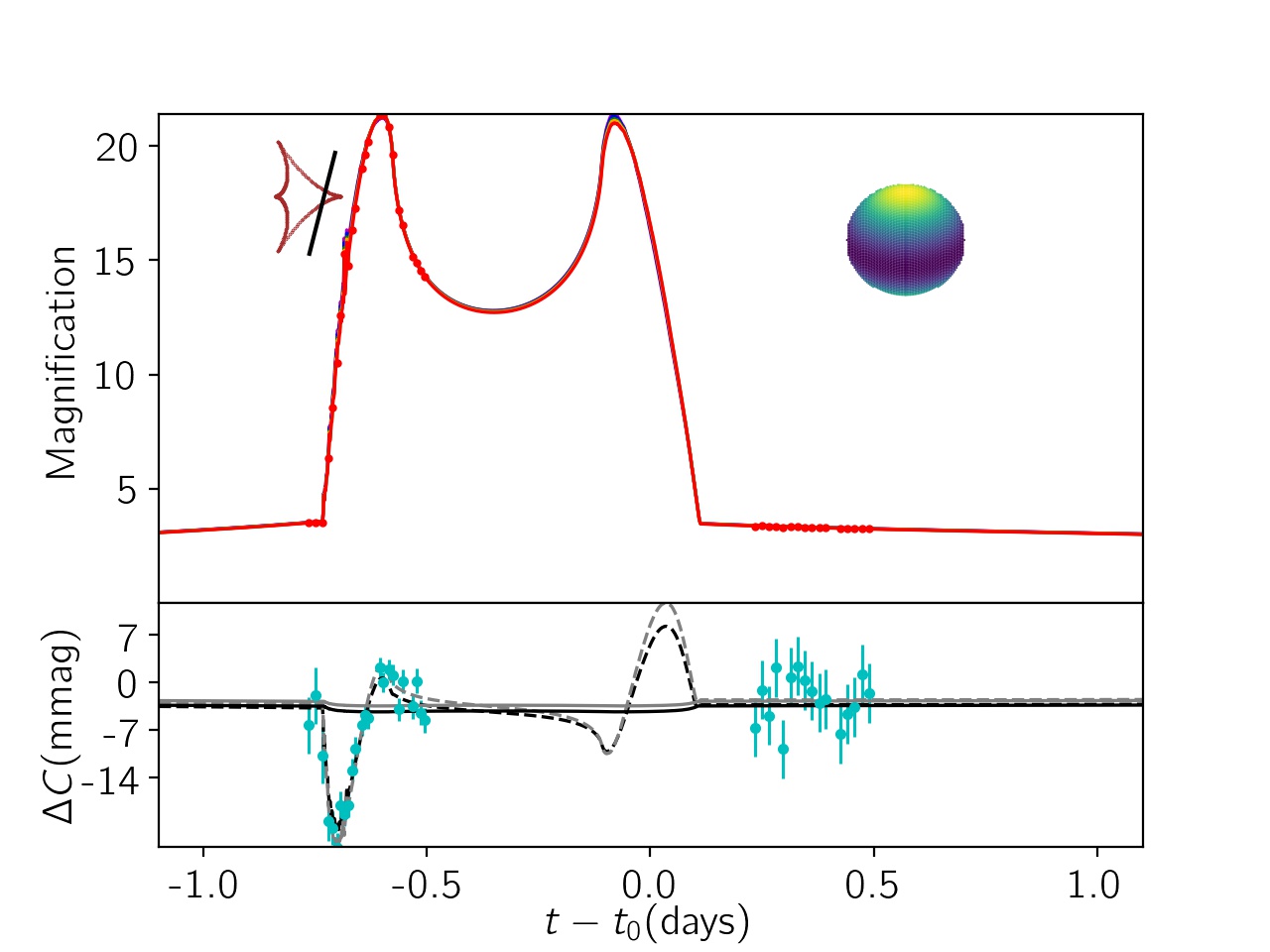}\label{fig4c}}
\subfigure[]{\includegraphics[angle=0,width=0.49\textwidth,clip=0]{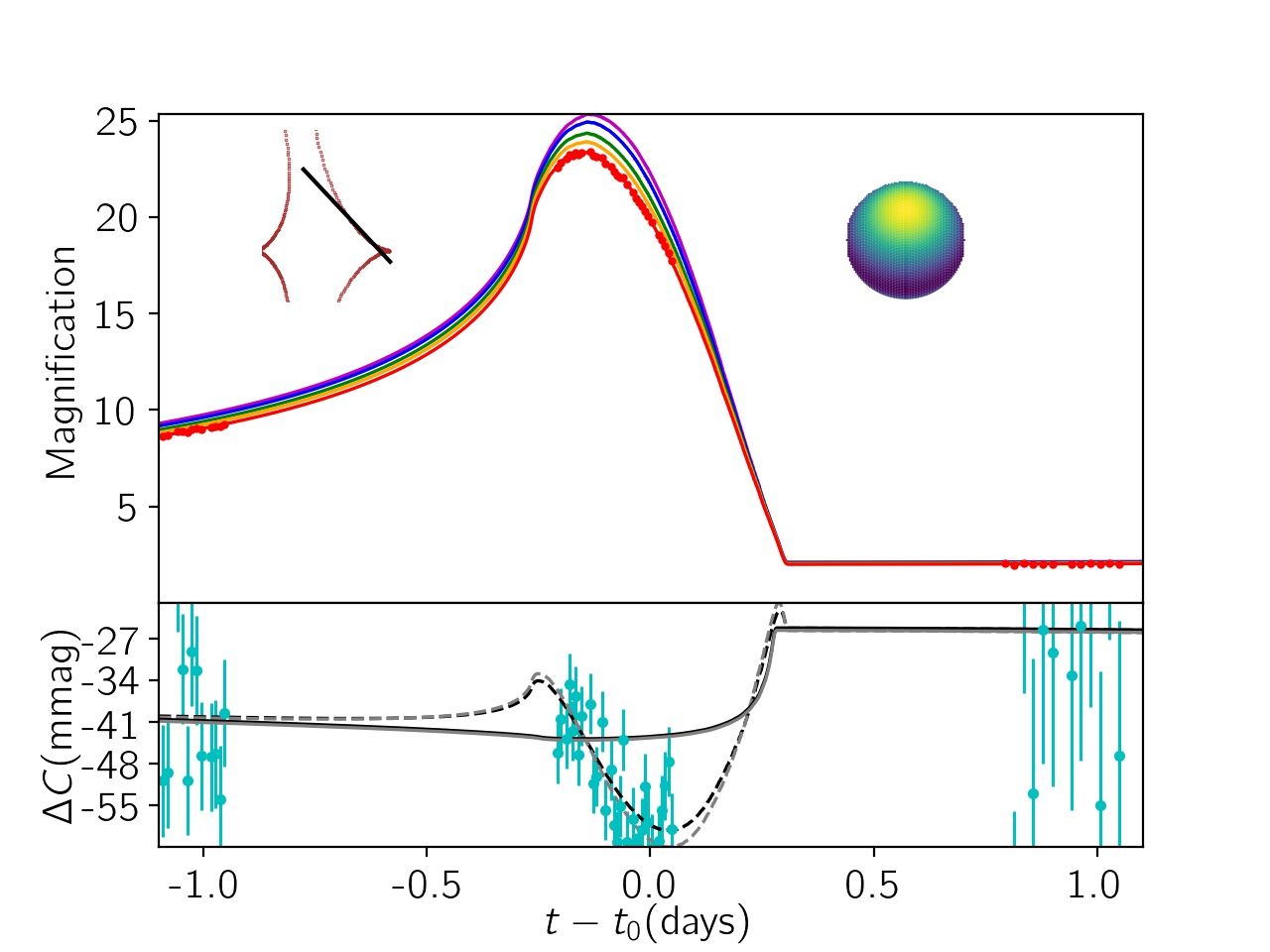}\label{fig4d}}
\caption{Similar to Figures \ref{three}, \ref{last} and \ref{four}, but for early-type source stars by considering the GD and BL effects. The parameters of the lightcurves are given in Table \ref{tab4}. The right-hand insets (colored schemes) represent brightness profiles over gravity-darkened source disks and the projected lens trajectory (black lines). In the residual parts, the perturbations in stellar colors due to the GD and BL effects are shown by dashed curves. The corresponding BL-induced chromatic perturbations are represented by solid curves.}\label{five}
\end{figure*}
\begin{table*}
\caption{The parameters of the lightcurves shown in Figure \ref{five}.}             
\label{tab4}      
\centering          
\begin{tabular}{ccccccccccccc}    
\toprule[1.5pt]
$~$&$\log_{10}[u_{0}]$&$t_{\rm{E}}$&$\log_{10}[\rho_{\ast}]$&$m_{\rm{base},~I}$&$T_{\rm{eff}}$&$\Omega$&$i$& $b_{I}$&$q$&$d$&$\xi$&$\Delta \chi^{2}$\\
&&$\rm{(days)}$&&$\rm{(mag)}$&$\rm{(K)}$&&$\rm{(deg)}$&&&&$\rm{(deg)}$&\\ 
\toprule[1.5pt]
\ref{fig4a}& -2.18 & 9.2 &  -2.09 &  18.0 & 7060 & 0.79 & 22.81& 0.99 & - & - & 255 & 33\\%% 26th Aug
\ref{fig4b}& -2.27 & 8.7 &  -1.95 &  16.3 & 7703 & 0.70 & 71.18& 0.92& - & - & 252 & 14757\\%%14   24th Aug
\ref{fig4c}&-0.85 & 10.5 &  -2.15 &  15.4 & 10199 & 0.67 & 27.58 & 0.99 & 0.26 & 0.68 & -104 & 675 \\%% 26th Aug
\ref{fig4d}&-0.70 & 21.6 & -2.69 &  17.3 & 7060 & 0.68 & 56.00 & 0.89 & 0.62 & 0.86 & -47 & 3302 \\% 26th Aug
\hline
\end{tabular}
\tablefoot{Here, $\Omega$ is the stellar angular velocity normalized to the critical velocity, and $i$ is the inclination angle of the stellar pole with respect to the sky plane.}
\end{table*}
\section{Gravity darkening}\label{gradark}

Early-type stars such as Be-stars usually rotate very fast and therefore have elliptical shapes. As a result their surface brightness is not uniform and it increases from the stellar equator toward the poles, the so-called gravity-darkening (GD) effect \citep[see, e.g., ][]{Lebovitz1967,Maeder1970}. 

Gravitational microlensing of elliptical source stars have been studied in several references. The first semi-analytical formalism for deriving the magnification factor of elliptical source stars was offered by \citet{Heyrovski1997}. The effect of ellipticity of the source stars in binary microlensing lightcurves and close to the caustic fold was also investigated by \citet{Gaudi2004, Alex2018}. The GD effect and elliptical shapes of early-type stars change the polarization curves of HM microlensing events and break its symmetry with respect to the time of the closest approach \citep{sajadian2016}.

The GD effect causes temperature variations over the source disk, such that the stellar temperature is highest at the poles and decreases toward the stellar equator. The temperature variations over the source surface depend on the effective surface gravity which is given by:  
\begin{eqnarray}\label{geff}
\bf{g}_{\rm{eff}}(\omega, \theta_{\ast}) &=& \Big[-\frac{GM_{\ast}}{R_{\ast}^{2}(\theta_{\ast})} + \omega^{2} R_{\ast}(\theta_{\ast}) \sin^{2} \theta_{\ast}\Big]\bf{e}_{r} \nonumber\\
&+&\Big[\omega^{2} R_{\ast}(\theta_{\ast}) \sin \theta_{\ast} \cos \theta_{\ast} \Big] \bf{e}_{\theta},
\end{eqnarray}
\noindent In this relation, $M_{\ast}$ is the mass of the source star, $\omega$ is the stellar angular velocity around its polar axis, and $\theta_{\ast}$ is the polar angle in the stellar coordinate system. $\bf{e}_{r}$ and $\bf{e}_{\theta}$ are the radial and polar unit vectors in the stellar coordinate system. $R_{\ast}$ is the stellar radius which for an ellipse depends on the polar angle $\theta_{\ast}$ in the following way \citep{Maeder2012}:
\begin{eqnarray}\label{rtet}
R_{\ast}(\theta_{\ast})= \frac{3 R_{\rm p}}{\Omega~\sin \theta_{\ast}} \cos \Big[\frac{\pi + \cos^{-1} (\Omega~\sin \theta_{\ast})}{3}\Big],
\end{eqnarray}
where, $R_{\rm p}$ is the stellar radius at its poles. $\Omega= \omega/ \omega_{\rm{c}}$ is the stellar angular velocity normalized to the critical (or break up) velocity, which is  $\omega_{\rm{c}}^{2}=(2/3)^{3}~G~M_{\ast}/R_{\rm p}^{3}$, i.e., the velocity at which the gravitational force at the stellar equator equals the centrifugal force.

\noindent The effective temperature of an elliptical source star which rotates around its polar axis depends on $\theta_{\ast}$ and can be given as \citep[see, e.g.,][]{VonZeipel,Chandrasekhar}:  
\begin{eqnarray}
T(\omega,\theta_{\ast})= T_{\rm{pole}}\Big[ ~\frac{g_{\rm{eff}}(\omega,\theta_{\ast})}{G~M_{\ast}/R_{\rm{p}}^{2}} ~ \Big]^{1/4},
\end{eqnarray}
where $g_{\rm{eff}}(\omega, \theta_{\ast})$ is the value of the effective gravity vector given by the equation \ref{geff}. $T_{\rm{pole}}$ is the surface temperature at the stellar poles.

In order to model an elliptical source star in the observer's coordinate system, we use a inclination angle $i$ which is the angle between the sky plane and the stellar polar axis. By rotating the source star around $x_{\ast}-$axis (which is horizontal and toward right) by the angle $-i$ one can convert the stellar coordinate system to the observer one and project the source star onto the plane of the sky. The projection process changes the source shape and its eccentricity which can be given by: 
\begin{eqnarray}
e= \sqrt{1- \frac{R^{2}_{\rm p}}{R_{\rm p}^{2} \cos^{2} i + R^{2}_{\rm{eq}} \sin^{2} i}}, 
\end{eqnarray}
where the source radius $R_{\rm{eq}}$ at the equator in the stellar coordinate system is given by Equation \ref{rtet} when setting $\theta_{\ast}=90^{\circ}$, such that:
$$R_{\rm{eq}}= \frac{3R_{\rm p}}{\Omega} \cos \big[ \frac{\pi+ \cos^{-1}\Omega}{3}  \big].$$
In the observer's coordinate system, the stellar pole is on the vertical axis on the sky plane and at the position $R_{\rm p} \cos i$ with respect to the source center. If $i=0$, the observer will see both stellar poles. 

After simulating an early-type source star, we turn on the lensing effect. For evaluating the magnification factor, we integrate the stellar intensity over the projected source surface. The intensity of each element of the source surface depends on its temperature and is determined by the corresponding Planck function. We simulate a large ensemble of HM microlensing events of elliptical source stars by considering the BL and GD effects. In this regard, we choose the inclination angle and the normalized angular velocity uniformly from the ranges $i \in [0,~90]$ deg and $\Omega \in [0.2,~0.9]$. The stellar temperature at its poles $T_{\rm{pole}}$ is calculated according to its mass and using the Besan\c{c}on model \citep{Robin2003, Robin2012}. 

\noindent Four examples of these lightcurves are shown in Figure \ref{five}. The brightness profile of the source star projected on the sky plane and the lens trajectory (solid black lines) are shown in the insets of each lightcurve panel. The parameters of these lightcurves are mentioned in Table \ref{tab4}. Some key points are listed here:

\begin{itemize}
\item  If the lens is passing over (or close to) the projected position of the stellar poles, the magnification in the $U,~B$-bands enhances, whereas if the lens is crossing the stellar equator the magnification factor increases in longer wavelengths, e.g. the $R, I$-bands. As a results, the chromatic perturbations are most often larger in $B$-$R$ than in $V$-$I$ (e.g., Figures \ref{fig4a} and \ref{fig4b}). This point can be inferred from two the rows (with the label GD) of  Table \ref{tab_result}.
	
\item The chromatic perturbations for these hot stars are asymmetric in time and include consecutive negative-positive deviations, unless either the lens trajectory is parallel with the stellar equator or $i=90^{\circ}$.
	
\item Unlike the chromatic perturbations due to the LD effect, the maximum amount of the GD-induced perturbations does not happen at the time of the closest approach (unless $i\sim 90$ deg). Their maxima occur when the lens is passing over (or close to) the stellar pole and its equator.   
	
\item According to the simulation, the amplitude in the chromatic perturbations are $\Delta C_{VI,~\rm{max}}\sim0.01,~0.03,~0.06$ mag for stars with $\Omega \simeq 0.5,~0.8,~0.9$, respectively. The perturbations will be best observable when the lens is passing very close to the stellar poles. One example lightcurve is shown in Figure \ref{fig4b}.
	
\item We note that the elliptical shape of the source star does not change its color while lensing and the gravity darkening effect does change it.   
	
\item If the lens does not cross the source surface, the amplitude of the chromatic perturbation is very small and less than $0.02$mag. However, the amplitude depends on $\Omega$.  

\item For early-type stars as source stars (which are intrinsically bright) the blending effect is very small.
\end{itemize}

By applying $\Delta \chi^{2}>100,~200$ as detectability criteria (with low and high sensitivities), we conclude that the resulting chromatic perturbations are detectable by the LI camera at the Danish telescope in $62,~54\%$ of HM microlensing events of early-type source stars (with $\Omega \in [0.2,~0.9]$). The main reason for such high detection efficiencies is the intrinsic brightness of early-type stars (see Table \ref{tab4}). During CC features, this probability is $41$-$48\%$. In these events, the chromatic signatures can reach to $0.4,~0.6$ mag for stars with $\Omega \simeq 0.7,~0.9$. On average, the maximum chromatic deviation in $B$-$R$ is $0.02$-$0.03$ mag.  Measuring chromatic perturbations while lensing will give us some information about the projected location of stellar pole on the sky. However, there is a degeneracy between parameters, because their number is larger than the number of observable quantities.

\section{Summary and Conclusions}\label{result}

The bending angle of light in a gravitational field does not depend on the wavelength of the light. If the brightness of the source star in a microlensing event is homogeneous over the surface and there is no blending effect, then the observed color of the source will not change during the lensing. One way to distinguish microlensing events from other astrophysical phenomena in which the stellar brightness changes with time is therefore to check whether the stellar color changes with the changing brightness. However, if the color of the source star is not identical all over its surface or the blending effect is considerable, the different magnification factors in different filters will cause the color of the lensing event to change as function of time. These chromatic variations are most pronounced in HM and binary CC microlensing events.

The surface brightness of the source stars may vary due to (i) the limb-darkening, (ii) close-in giant planet companions, (iii) stellar spots and (iv) gravity-darkening. In all of these situations, the stellar color changes as function of time during the lensing. Additionally, the BL effect  in microlensing events makes chromatic perturbations during lensing. In fact, the blending parameter depends on the adopted filter. In this work, we have evaluated the resulting perturbations in the $V$-$I$ and $B$-$R$ stellar colors in HM and CC events by considering the aforementioned effects. Since, HM and CC microlensing events are followed by use of the dual-color LI camera on Danish $1.54$ m telescope by our team, we have in the present work estimated the probability of detecting these perturbations in the $V$-$I$ stellar color by use of this camera which were summarized in Table \ref{tab_result}.\\

The importance of studying the chromatic perturbations are twofold. (I) Since the perturbations are time-dependent, then different sets of data points taken for a microlensing event in different filters (often by different telescopes) cannot be converted to the same filter in order to determine a best-fitting model. The relations for converting magnitudes in different filters are applicable only when the lens-source distance is sufficiently large, such than the source radius does not give rise to a finite-source effect (i.e. near the baseline of the event). Ignoring these perturbations while modeling HM or CC microlensing events creates extra noise in the converted data points (especially the data around the magnification peaks). (II) Dual-color instruments can potentially measure these chromatic perturbations and allow determination of the variation in the surface brightness and color of the source stars. That is another channel for probing the surface of Galactic bulge stars. 

%%%%% blending 
The BL effect in microlensing events makes chromatic variations for the source stars. The movement of source stars in the color-color diagram (due to the BL-induced chromatic perturbations) helps to determine the stellar types \citep{Tsapras2019}. The BL-induced chromatic perturbation enhances by increasing the magnification peak. Its increasing slope tends to zero for the magnification factor larger than $40$. The time scale of these chromatic perturbations is of order $t_{\rm E}$ (the time scale of the lensing effect). For transit microlensing events (with the considerable finite-source effect), the magnification factor remains almost constant while passing the source surface. Hence, in these events and around their peaks the BL-induced chromatic perturbations are almost constant and can be subtracted from the chromatic variation curves.

%%%%%% LIMB-DARKENING
By simulating HM microlensing events of source stars and including LD effect, we concluded that for transit microlensing events, the lightcurves are more flattened in longer wavelength (lower peak and larger width) than at shorter wavelengths. In these events, the chromatic perturbations maximize at the time of the closest approach and can even reach $0.04$mag (which measurable by our LI camera). For non-transit microlensing events, the peaks of lightcurves maximizes in the $I$-band and the maximum chromatic perturbations reach $0.01$ mag. During CC and when the source edge is passing from the caustic line and the source center is out of the caustic, these chromatic signals are higher and reaches $0.07$ mag. We note that the chromatic perturbations due to LD effects in the $B$-$R$ color is higher than in $V$-$I$. The probabilities of detecting LD and BD-induced perturbations in $V$-$I$ by our LI camera are $50$-$60$ ,~$20$-$30\%$ in HM and CC events, respectively. We considered $\Delta \chi^{2}>200,~100$, as the detectability threshold. 

%%%%%%%%%%%%%%%%%%% CGPS
Close-in giant planets orbiting the source stars with inclination angle $\sim 90$ deg (i.e., edge-on orbits), transit their host stars and make periodic changes in their brightness and color distribution over the surface. We studied if this kind of variation in the stellar brightness can make significant chromatic perturbations during lensing. The probability for detecting planet-induced chromatic perturbations for face-on planetary orbits is low, because it would require that the lensing star passes very close to (or over) the planet surface, which is small. For edge-on orbits the probability is higher, because it is the change in color of the source star (where the planet transits) that is amplified by the lens, and causes the achromatic effects, and reveals the planet. 
By including both the LD, BL and CGPs orbiting the source stars, we concluded that the probabilities for detecting the chromatic perturbation in HM or CC events are $53$-$64$,~ $23$-$32\%$, respectively. We note that CGPs improve the detection probability
of chromatic perturbations by $\sim 2\%$.

%%%%%%%%%STELLAR SPOTS
By adding the stellar spots for limb-darkened and blended source stars, these probabilities increases to $51$-$63$,~ $27$-$36\%$ in HM and CC events, respectively. In these events, if the lens transits the stellar spot, an extra bump appears in the $V$-$I$ light curve of the source star which alters the stellar color up to $0.05$ mag. The spot-induced perturbations in the $V$-$I$ color are higher than in $B$-$R$. The SS-induced chromatic perturbations  can be degenerate with those made by transiting CGPs.

%%%%%%%%%%%%% GRAVITY DARKENING
Early-type source stars do not have constant surface brightness because of the gravity-darkening effect which is resulting from the elliptical shape of these stars. If the lens transits the source surface and passes close to (or over) the stellar poles, the chromatic perturbation maximizes and reaches $0.01,~0.03,~0.06$ mag for stars with $\Omega \sim 0.5,~0.8,~0.9$, respectively. Our LI camera can potentially discover these perturbations with a probability of $54$-$62$,~$41$-$48\%$ if it observes HM and CC events.  

%%%%%%%%%%%I%%%%%%%%%%%%%%%%%%%%%%%%%%%%%%%%%%%%%%%%%%%%%%%%%%%%%%%%%%%%%%
\begin{acknowledgements}
We acknowledge support from the European Union H2020-MSCA-ITN-2019 under Grant no. 860470 (CHAMELEON) and from the Novo Nordisk Foundation Interdisciplinary Synergy Program grant no. NNF19OC0057374. S. Sajadian thanks the Department of Physics, Chungbuk National University and especially C.~Han for hospitality. We thank M.~Hundertmark, J.~Scottfelt and D.~Bramich for consultations. We appreciate the referee for the helpful comments and good suggestions. 
\end{acknowledgements}

\bibliographystyle{aa} % style aa.bst
\bibliography{references}

\clearpage
\newpage
\begin{appendix}  \label{append}
\section{Photometric uncertainties of data taken by the Danish  telescope}
\begin{figure}
\centering
\includegraphics[angle=0,width=0.49\textwidth,clip=]{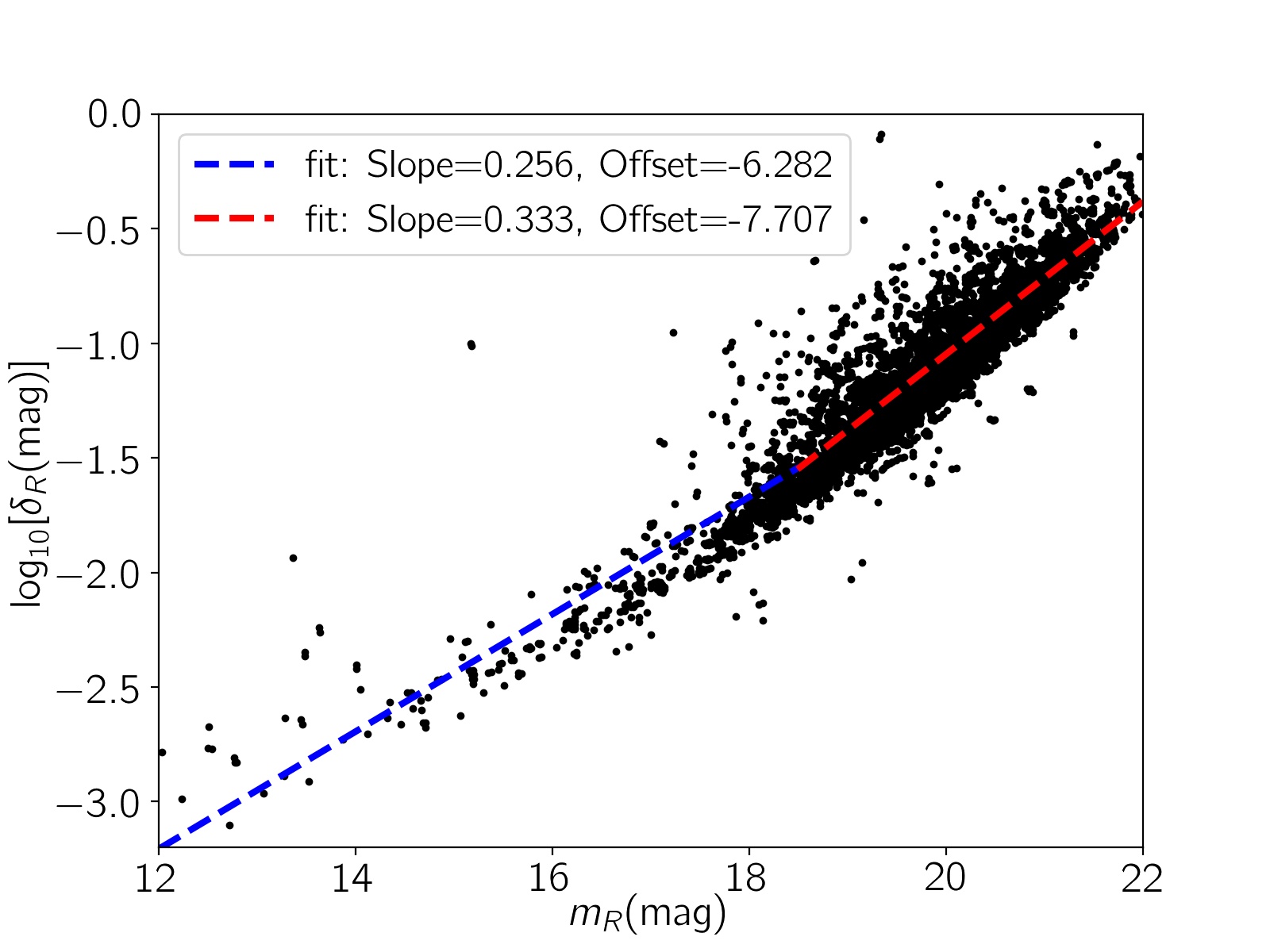}
\caption{The scatter plot of RMS magnitude deviation versus mean $R$-band magnitude as measured by the LI camera on Danish telescope and reported in \citet{Scottfelt2015}. The dashed lines represent the best-fitted linear curves to black data points and given by Equations \ref{fittre}.}\label{errordd}
\end{figure}
\begin{table}
\caption{The best-fitted parameters of linear relation between $m_{V}+m_{I}$ versus $m_{R}$ for different values of the stellar age and metalicity. A complete electronic version of this table is available at: .... }             
\label{tablast}      
\centering          
\begin{tabular}{c c c c}    
\toprule[1.5pt]
$[\rm{Fe}/ \rm{H}]$&$\rm{Age(Gyr)}$&$\rm{b(mag)}$&$\rm{a(mag)}$\\
\toprule[1.5pt]
		-2.51 & 0.25 & -0.0044 $\pm$0.00054 & 2.0036 $\pm$0.00018\\
		-2.51 & 0.30 & -0.0065 $\pm$0.00053 &  2.0038 $\pm$0.00018\\
	    -2.51 & 0.35 & -0.0089 $\pm$0.00053 &  2.0042 $\pm$0.00018\\
		-2.51 & 0.40 & -0.0099 $\pm$0.00052 &  2.0044 $\pm$0.00018\\
		-2.51 & 0.45 & -0.0156 $\pm$0.00065 &  2.0060 $\pm$0.00022\\
		-2.51 & 0.50 & -0.0181 $\pm$0.00062 &  2.0061 $\pm$0.00021\\
		-2.51 & 0.55 & -0.0205 $\pm$0.00061 &  2.0060 $\pm$0.00021\\
		-2.51 & 0.60 & -0.0195 $\pm$0.00084 &  2.0054 $\pm$0.00024\\
		-2.51 & 0.65 & -0.0163 $\pm$0.00087 &  2.0043 $\pm$0.00025\\
		-2.51 & 0.70 & -0.0143 $\pm$0.00075 &  2.0039 $\pm$0.00024\\
		$......$ \\
		\hline
	\end{tabular}
\end{table}

The error bars of data taken by LI camera on the Danish telescope in the Johnson $R$-band were well studied in \citet{Scottfelt2015}.  The authors have measured the root mean square (RMS) magnitude deviations, $\delta_{R}$, versus the stellar apparent magnitudes in the $R$-band. The data-reduction of their images have been done using \texttt{DanIDL}\footnote{\url{http://www.danidl.co.uk/}} pipeline \citep{Bramich2008, Bramich2013}. We use their results to determine the error bars of synthetic data points taken by LI camera in our simulation. We show their measurements in Figure \ref{errordd} and fit a hybrid linear curve to the data. The equation of the fitted hybrid function is: 
\begin{eqnarray}\label{fittre}
\log_{10}[\delta_{R}]=\nonumber~~~~~~~~~~~~~~~~~~~~~~~~~~~~~~~~~~~~~~~~~~~~~~~~~~~~~~~~~~~~~~~~~ \\
\begin{cases}
0.256(\pm 0.005)~m_{R} -6.282(\pm 0.09) & m_{R} \leq 18.5\\
0.333(\pm 0.004)~m_{R} -7.707(\pm 0.09) &  m_{R}>18.5.  
\end{cases}
\end{eqnarray}
In order to estimate error bars in standard Johnson filters $V$ and $I$ for synthetic data taken by LI, we use two relations between the photometric measurements in $R$ and $V$, $I$-bands. These relations are explained in the following.

The stellar magnitude in the $R$-band is approximately the average of the corresponding magnitudes in the $V$ and $I$-bands. However, the stellar age and metalicity change the stellar spectra somewhat. In order to find an accurate relation between the stellar magnitudes for any given values of stellar age and matelicity, we use the Dartmouth stellar Isochrones \footnote{\url{http://stellar.dartmouth.edu/models/}} \citep{2008Dotter}. For any given values of stellar age and matellicity and for different stars with different initial mass, we fit a linear relation between the $R$-band and $V+I$-band magnitudes, as $m_{V}+m_{I}=  a~m_{R} + b$. In Table \ref{tablast}, the slope and offset of these linear relations for different stellar age and metalicity values are given.
	
This relation results:
\begin{eqnarray}\label{rel1}
\sqrt{\delta_{V}^{2}+ \delta_{I}^{2} } = a~\delta_{R}. 
\end{eqnarray}

Another relation between these error bars is necessary to determine their contributions. In our simulation and for each given value of magnification factor, $A_{F}(t)$, we can determine the ratio of signal to noise ratio (SNR) in these two filters as following: 
\begin{eqnarray}\label{snr}
\eta(t)&=&\frac{\rm{SNR}_{V}}{\rm{SNR}_{I}} =\sqrt{\frac{10^{-0.4 m_{V}} A_{V}(t)~+~10^{-0.4 \mu_{\rm{sky},V}}\Omega_{V}~+~\mathcal{F}_{b,~V}}{10^{-0.4 m_{I}} A_{I}(t)~+~10^{-0.4 \mu_{\rm{sky},I}}\Omega_{I}~+~\mathcal{F}_{b,~I}}}\nonumber\\
&\times& 10^{0.2 \left(m_{\rm{zp},V}-m_{\rm{zp},I}\right)},
\end{eqnarray}

\noindent where, $\Omega=\pi (\rm{FWHM}/2)^{2}$ is the area of the stellar PSF.  FWHM is the Full Width at Half Maximum of the stellar brightness profile. For LI camera on the Danish telescope $\rm{FWHM} \simeq 0.48,~0.46$ arcs in $V$ and $I$-bands  \citep{Scottfelt2015}. The sky brightness at the La Silla observatory was reported as $\mu_{\rm{sky}}=21.75,~19.48~\rm{mag.~arcs^{-2}}$ in $V$ and $I$-bands, respectively \citep{Mattila1996}. We do not need to include the exposure time in this ratio, because LI camera has the same exposure times for images taken in these two filters. This ratio, $\eta(t)$, gives another relation between the photometric error bars in $V$ and $I$-bands as following:  

\begin{eqnarray}\label{snr2}
\eta= \frac{\left|10^{0.4 \delta_{I}}-1 \right| }{\left|10^{0.4 \delta_{V}} -1 \right|}.  
\end{eqnarray}

\noindent Using the relations \ref{rel1} and \ref{snr2}, we determine photometric error bars for each couple of data points taken by LI camera on the Danish telescope in $V$ and $I$-bands. We note that a similar method was used to estimate the error bars of data points taken by the Danish telescope in \citet{Evans2018,Evans2016}.
\end{appendix}
\end{document}